\documentclass[11pt]{article}

\usepackage[left=1in,top=1in,right=1in,bottom=1in,head=.1in,nofoot]{geometry}

\usepackage{setspace,url,bm,amsmath} % For double-spacing, URL font, math symbols

\usepackage{titlesec} % Section header formatting
\titlelabel{\thetitle.\quad} % Section header formatting
\usepackage[margin=20pt]{subcaption}

\usepackage{amsmath,amsfonts,amsthm,amssymb}
\usepackage{graphicx}

\usepackage[colorlinks=true,linkcolor=black,citecolor=blue,urlcolor=blue]{hyperref}
\usepackage{threeparttable}
\usepackage{manyfoot}%
\usepackage{booktabs}%
\usepackage{algorithm}%
\usepackage{algorithmicx}%
\usepackage{algpseudocode}%
\usepackage{listings}%
\usepackage{dsfont}%
\usepackage{bm}
\usepackage{etoolbox}
\usepackage{natbib}
\usepackage{adjustbox}
\usepackage{threeparttable}
\usepackage{dcolumn}
\usepackage{subcaption}
\theoremstyle{definition}

\DeclareFontFamily{U}{mathx}{\hyphenchar\font45}
\DeclareFontShape{U}{mathx}{m}{n}{
      <5> <6> <7> <8> <9> <10>
      <10.95> <12> <14.4> <17.28> <20.74> <24.88>
      mathx10
      }{}
\DeclareSymbolFont{mathx}{U}{mathx}{m}{n}
\DeclareFontSubstitution{U}{mathx}{m}{n}
\DeclareMathAccent{\widecheck}{0}{mathx}{"71}
\DeclareMathAccent{\wideparen}{0}{mathx}{"75}

\allowdisplaybreaks[4]

\newtheorem{assump}{Assumption}
\newtheorem{thm}{Theorem}
\newtheorem{lemma}{Lemma}
\newtheorem{remark}{Remark}

\newtheorem{prop}{Proposition}

\newcommand{\norm}[2]{\lVert#1\rVert_#2}
\newcommand{\source}[1]{#1^{(s)}}

\newcommand{\ProjBeta}{\beta_{[k]\rm proj}(a)}
\newcommand{\ProjBetaSource}{\beta^{(s)}_{[k]\rm proj}(a)}
\newcommand{\ProjBetaSourceE}{\hat{\beta}^{(s)}_{[k]\rm proj}(a)}
\newcommand{\HLassoBeta}{\hat\beta_{[k]\rm lasso}(a)}
\newcommand{\diag}{\mathrm{diag}}
\newcommand{\var}{\mathrm{var}}
\newcommand{\fix}[1]{\textcolor{black}{#1}}
\newcommand{\del}[1]{\textcolor{black}{#1}}

\title{\vspace{-1.5cm}Incorporating external data for analyzing randomized clinical trials: A transfer learning approach}
\author{
\small
{
Yujia Gu$^{1}$, \ \ Hanzhong Liu$^{2}$, \ \ Wei Ma$^{1}$\thanks{\small{Correspondence: \texttt{mawei@ruc.edu.cn}}}
}
\\
{\small $^{1}$ Institute of Statistics and Big Data, Renmin University of China, Beijing, China}\\
{\small $^{2}$ Center for Statistical Science, Department of Industrial Engineering, Tsinghua University, Beijing, China}
}
\date{}

\begin{document}
\doublespacing

\maketitle
\vspace{-1cm}
\begin{abstract}

Randomized clinical trials are the gold standard for analyzing treatment effects. However, increasing costs and ethical concerns may limit trial recruitment, resulting in insufficient sample sizes and potentially invalid inference. Incorporating external trial data with similar characteristics (treatments, diseases, biomarkers, etc.) into the analysis appears promising for addressing these issues. Transfer learning, which in our context utilizes external trials as the source domain and current trials as the target domain, may offer a viable approach. In this paper, we present a formal framework for applying transfer learning to the analysis of clinical trials, considering three key perspectives: transfer algorithm, theoretical foundation, and inference method. To cover broad types of randomized trials, we study this problem under stratified, or more generally, covariate-adaptive randomization. For the algorithm, we adopt a parameter-based transfer learning approach to enhance the lasso-adjusted stratum-specific estimator developed for estimating the treatment effect. A key component in constructing the transfer learning estimator is deriving the regression coefficient estimation within each stratum, accounting for the bias between source and target data. To provide a theoretical foundation, we derive the $l_1$ convergence rate for the estimated regression coefficients and subsequently establish the asymptotic normality for the transfer learning estimator. Our results show that when external trial data resembles current trial data, the sample size requirements can be reduced compared to using only current trial data. Finally, we propose a consistent nonparametric variance estimator to facilitate inference that is robust to model misspecifications and applicable to various commonly used randomization procedures. Numerical studies demonstrate the effectiveness and robustness of our proposed estimator across various scenarios. Our study highlights the potential of transfer learning in analyzing randomized clinical trials.

\vspace{10pt}

\end{abstract}

\section{Introduction}

%简单的RCT + lack of units 

%Randomization is considered a gold standard in clinical trials and other intervention studies, such as online A/B tests and other experiments in economics. After randomization, we aim to find a treatment effect estimator that can provide valid inference. However, randomized clinical trials may suffer from rising costs, ethical problems, and difficulties in volunteer recruitment, for example, when the trial is for some rare or life-threatening diseases. As a consequence, the inference becomes invalid due to the lack of units. To solve this problem, data from external clinical trials are encouraged to assist in analyzing the current trial \citep{FDA2022}; for discussions on using external data from observation studies, see related work in Section \ref{sec1.1}. Especially, the external trial data can be used when the current trial and external trial share similar drugs, biomarkers, or treatment procedures. For example,  researchers leveraged external Recovery trial data to evaluate the efficacy of high-dose dexamethasone in critically ill COVID-19 patients to empower the inference, although the Bayesian methods they used may be sensitive to the parameter settings \citep{chevret2022challenges}. In this paper, we aim to propose an integration approach that takes advantage from leveraging external trial data while still providing valid inferences that are robust against model misspecification.

Randomization is the gold standard in clinical trials and other intervention studies, such as online A/B tests and experiments in development economics. After randomization, a treatment effect estimator that can provide valid inference should be constructed. However, randomized clinical trials may suffer from rising costs, ethical concerns, and recruitment difficulties, especially when the trial involves rare or life-threatening diseases. Consequently, the inference may become invalid due to the lack of sufficient experimental units, i.e., patients enrolled in the trial. To solve this problem, data from external trials are encouraged to assist in analyzing the current trial \citep{FDA2022}; for discussions on using external data from observational studies, see related work in Section \ref{sec1.1}. The use of external trial data is most promising when the current trial and the external trial share similar drugs, biomarkers, or treatment procedures. For example, when analyzing the Covidicus trial, researchers leveraged external Recovery trial data to evaluate the efficacy of dexamethasone, a common treatment in both trials, for critically ill COVID-19 patients, although the Bayesian methods used may be sensitive to parametric model settings \citep{chevret2022challenges}.

The recent development of transfer learning, especially in the field of statistics, offers insights into incorporating external trial data into the analysis of current trial data. Transfer learning \citep{torrey2010transfer} uses information from similar source data to support the analysis of target data. To date, transfer learning has been successfully applied in many fields, such as natural language processing, computer vision, and epidemiology. Due to its empirical success, transfer learning has received considerable attention in statistical disciplines. For example, it has been studied in a variety of problems, such as classification \citep{cai2021transfer}, nonparametric regression \citep{cai2024transfer-nonpara}, and contextual multi-armed bandits \citep{cai2024transfer-context}. Most relevant to our work, due to the availability of a large number of baseline covariates in modern clinical trials, transfer learning, when applied to high-dimensional problems, has been shown to improve the convergence rate of regression coefficients compared to using only the target data \citep{Li2022Transfer,tian2023transfer}. Moreover, transfer learning can also improve estimation and inference accuracy in the Gaussian graphical model \citep{li2022GGM}. These works inspire us to apply transfer learning to estimating treatment effects in clinical trials, especially in a high-dimensional setting.

While incorporating external trial data through transfer learning is conceptually intuitive, ensuring robust inference is pivotal for its acceptance in clinical trial practice, and this is the mission of our paper. Here, robustness indicates that the inference remains valid even when the working model used to derive the treatment effect estimation is misspecified. Such robustness is desirable when analyzing clinical trial data because the true data generation model is usually unknown. Without using external trial data, the robustness of treatment effect estimators has been extensively studied under various randomization methods. Specifically, under simple randomization, the robustness of different regression estimators has been discussed in several influential papers \citep{Freedman2008, lin2013, Yang2001}. In this paper, we focus on covariate-adaptive randomization, which is commonly used in clinical trials to produce a more balanced treatment allocation within strata formed by covariates, such as stratified block randomization \citep{zelen1974randomization}. According to recent surveys, covariate-adaptive randomization is implemented in about 70\% of trials \citep{ciolino2019ideal}. The inference under covariate-adaptive randomization has been challenging due to the dependence between treatment assignments. \citet{Bugni2018} proposed robust treatment effect estimators based on several regression models adjusting for strata variables. Subsequently, many researchers have adjusted additional baseline covariates in the regression models to further improve efficiency \citep{Ma2020regression, liu2022lasso, Ye2022jasa, Ye2022Biometrika, gu2023regression}. Notably, under a high-dimensional setting, \citet{liu2022lasso} applied lasso regression to obtain a lasso-adjusted stratum-specific treatment effect estimator, which is optimal among the class of all regression-adjusted estimators. For more discussion on covariate-adaptive randomization and its inference, please refer to Section \ref{sec1.1}.

% Problems
%Several challenges remain to be tackled. First, proposing a transfer learning approach is demanding due to the complexities involved in determining what knowledge or features should be transferred from the source to the target data and how to effectively transfer this knowledge without degrading the performance on the target task. Another challenge is providing theoretical guarantees that the treatment effect estimation can benefit from transfer learning, as the experiment units are dependent on the treatment design under covariate-adaptive randomization. Finally, valid inference is critical for treatment effect estimation, which is rarely considered in previous transfer learning research. We expect our transfer learning approach to produce inference that is applicable to commonly used randomization methods and robust against model misspecification.

To fulfill our mission, challenges from three perspectives need to be tackled: \textit{transfer algorithm}, \textit{theoretical foundation}, and \textit{inference method}. Firstly, a transfer learning approach must be tailored to the context of covariate-adaptive randomization to ensure proper knowledge or feature transfer from the source (external trial) to the target (current trial) data, without compromising the performance on the target task in terms of the robustness of treatment effect estimation. Secondly, providing theoretical guarantees that the treatment effect estimation will benefit from transfer learning is crucial, given that \fix{the dependence in treatment assignments across experimental units resulting from covariate-adaptive randomization}. Finally, valid inference is critical for accurate treatment effect estimation, which is rarely considered in previous transfer learning research. We expect our transfer learning approach to produce valid inference that is applicable to commonly used randomization methods and robust against model misspecification.

%In this paper, we propose a robust treatment effect estimator via transfer learning to address the above issues. First, motivated by \citet{Li2022Transfer}, we apply a parameter-based transfer learning approach to treatment effect estimation under covariate-adaptive randomization. We begin by using lasso regression to obtain regression coefficients from the source data. Next, we correct the bias in regression coefficients between the source and target data using lasso regression on the target data. The target regression coefficient estimator is obtained by adding the source regression coefficient estimator and the bias estimator. Finally, the transfer learning estimator is constructed by replacing the regression coefficients in the stratum-specific estimator. 

In this paper, we propose a robust treatment effect estimator via transfer learning to address the aforementioned issues. For the first challenge, we apply a parameter-based transfer learning approach \citep{Li2022Transfer} to enhance the lasso-adjusted stratum-specific treatment effect estimator developed for covariate-adaptive randomization. Importantly, this method also strengthens data privacy and security, which is crucial in clinical trials due to the sensitivity of patient information. Specifically, we begin by using lasso regression to obtain regression coefficients from the source data within each stratum. We then correct the bias in these regression coefficients between the source and target data by applying lasso regression to the target data. The target regression coefficient estimator is obtained by adding the bias estimator to the source regression coefficient estimator. Finally, the transfer learning-enhanced treatment effect estimator (referred to hereafter as the transfer learning estimator for simplicity) is constructed by substituting the original regression coefficients in the stratum-specific estimator with these bias-corrected target regression coefficients.

%Then, building upon the robust lasso-adjusted treatment effect estimator in \citet{liu2022lasso}, we establish the $l_1$ norm consistency of the target regression coefficient estimators, which is the key to deriving the asymptotic normality for the transfer learning estimator. When the source and target data are similar under mild conditions, the $l_1$ convergence rate of the regression estimators is faster than the convergence rate in \citet{liu2022lasso}. This result leads to an improvement in the asymptotic normality of our proposed estimator, which requires fewer units in target data. Notably, our proposed estimator obtains the same efficiency as the stratum-specific estimator, identified as optimal in work by \citet{gu2023regression}. 

Then, extending the theory for the lasso-adjusted treatment effect estimator in \citet{liu2022lasso}, we establish the $l_1$ norm consistency of the target regression coefficient estimators, which is key to deriving the asymptotic normality for the transfer learning estimator. Under mild conditions, we show that, when the source and target data are similar, the $l_1$ convergence rate of the regression estimators is faster than the convergence rate in \citet{liu2022lasso}. This result leads to an improvement in inference based on our proposed estimator, which requires fewer units in the target data. Notably, our proposed estimator achieves optimal efficiency for a regression-adjusted estimator under covariate-adaptive randomization with multiple treatments \citep{gu2023regression}.

%Finally, we establish valid inference by developing a consistent nonparametric variance estimator for the transfer learning estimator. The inference is non-trivial under transfer learning for high-dimensional data. However, thanks to our framework, we can provide robust inference against model misspecification. Especially, the robustness of our transfer learning approach can adapt to more scenarios, where the covariate distributions for the source data and target data can be different, and the underlying model between potential outcomes and baseline covariates for the source data and target data can be distinct. 

Finally, we facilitate valid inference by developing a consistent nonparametric variance estimator for the transfer learning estimator. In general, inference under transfer learning for high-dimensional data is non-trivial. However, thanks to our framework, we can preserve robust inference for the treatment effect, avoiding negative transfer when incorporating source data. Moreover, the proposed inference procedure is widely applicable to a broad class of covariate-adaptive randomization procedures. Additionally, the robustness of our transfer learning approach can adapt to scenarios where the covariate distributions or functional forms of data generation models may differ between the source and target data.

\subsection{Related works}\label{sec1.1}

% RCT + CAR
\textit{Covariate-adaptive randomization and its inference}. Covariate-adaptive randomization \fix{aims to provide more balanced treatment allocation with respect to baseline covariates}. For example, stratified block randomization \citep{zelen1974randomization} defines sets of strata based on covariates and allocates experimental units in each stratum using block randomization. Minimization methods \citep{Taves1974, Pocock1975} have been proposed to balance covariates over their margins. This approach has been generalized to control various types of imbalance measures, including overall and within-stratum imbalance measures \citep{Hu2012,hu2020theory}. Other commonly used covariate-adaptive randomization approaches such as stratified biased coin design \citep{Efron1971, Shao2010} and model-based approaches \citep{Begg1980, Atkinson1982}. We refer readers to \citet{Rosenberger2015} for a detailed discussion of these methods.

% Inference
Our theoretical analysis of the inference for the regression-adjusted treatment effect estimators is related to recent studies, which have shown that under covariate-adaptive randomization, regression-adjusted treatment effect estimators are valid for inference. \fix{\citet{Bugni2018} proposed robust treatment effect estimators that are resistant to model misspecification with regression adjustment for stratification, which can produce valid inference.} With the inclusion of additional baseline covariates, stratum-common and stratum-specific treatment effect estimators have been proposed to improve the efficiency of the estimators \citep{Ye2022jasa, Ye2022Biometrika, Ma2020regression}. \citet{gu2023regression} showed that the stratum-specific estimator could gain efficiency compared with the stratum-common estimator under multiple treatments, regardless of whether the allocation ratios are the same across strata. In high-dimensional settings, \citet{liu2022lasso} obtained the regression coefficient estimator by lasso and replaced the original coefficient estimator in stratum-common and stratum-specific estimators to get lasso-adjusted estimators. Under regular conditions on sparsity, they showed that the lasso-adjusted estimators could achieve the same efficiency as the regression-adjusted estimators under two treatments. When the working model is a general function of covariates instead of a linear form, researchers reach a theoretical guarantee for the efficiency gain if the function form is properly estimated \citep{rafi2023efficient,bannick2023general,tu2023unified}. 
% CAR

\textit{Transfer learning}. Transfer learning is widely applied in many fields, such as computer vision \citep{wah2011caltech,saenko2010adapting,donahue2014decaf}, natural language processing \citep{ruder2019transfer,wolf2020transformers}, biomedical analysis \citep{schweikert2008empirical,petegrosso2017transfer}, etc. We refer readers to \citet{pan2009survey,weiss2016survey,zhuang2020comprehensive} for comprehensive surveys on transfer learning. Based on the type of knowledge being transferred, transfer learning problems can be categorized into several subproblems: \fix{instance-based, feature-based, parameter-based and relation-based problems} \citep{pan2009survey}. Our paper focuses on the parameter-based problem as we transfer the regression coefficients from the source data to the target data.
Due to its successful empirical performance, transfer learning draws extensive attention in the statistics field. Besides the work we have mentioned, transfer learning has also been applied to statistical problems such as high-dimensional quantile regression \citep{huang2022estimation,jin2024transfer} and precision medicine \citep{wu2023transfer}.

\textit{Bayesian historical borrowing}. In many cases, historical trial data with settings similar to the current trial data are available. Including these historical data may improve the precision of estimations in the current trial \citep{pocock1976combination}. As the treatments usually differ between trials, researchers mainly focused on using historical control data. However, the difference in population distribution may lead to heterogeneity among the historical data and current data. Under the Bayesian framework, researchers proposed several methods to address the issue, such as using power prior \citep{ibrahim2000power,ibrahim2015power}, meta-analytic predictive prior \citep{schmidli2014robust} and hierarchical models \citep{neuenschwander2010summarizing,hobbs2011hierarchical}. However, \del{almost all existing} Bayesian methods depend on the choice of prior distributions and the parametric model assumptions, potentially introducing bias and subjectivity into the statistical inference.  Also, previous studies only used historical control data, ignoring the potential improvement from using historical treatment data. 

\textit{Combining clinical trial and observation study}. Another choice of external data is real-world data, which mostly comes from observational studies, such as electronic health records, disease registry databases, wearable devices, etc. Observational studies have a much larger number of patients than clinical trials, which can empower the inference. However, it has disadvantages such as selection bias and unmeasured confounding, which may lead to biased estimates \citep{colnet2024causal}. Therefore, researchers have focused on effectively combining observational studies with clinical trials to leverage their respective strengths and compensate for weaknesses. For example, the integration can enhance the generalizability of clinical trial \citep{stuart2011use,dahabreh2020extending,lee2023improving}, improve estimation efficiency of treatment effect heterogeneity \citep{kallus2018removing,yang2023elastic,yang2020analysis, wu2022integrative}, and identify optimal individual regimes \citep{wu2023transfer,chu2023targeted}. \fix{For more discussion on combining randomized trials and observational studies, please refer to \citet{colnet2024causal}}.

% Historical control

% Transfer learning application

% Why different

\subsection{Paper outline}

The remainder of this paper is organized as follows. Section 2 introduces the framework and notations \fix{for} covariate-adaptive randomization. Before the main results, we generalize the lasso-adjusted treatment effect estimator for multiple treatments in Section 3. In Section 4, we develop our transfer learning algorithm and the asymptotic properties of the proposed transfer learning estimator. We propose nonparametric variance estimators for valid inference in Section 5. Simulation results are in Section 6. Section 7 provides a clinical trial example, and Section 8 summarizes our discussion. 

\section{Framework and notations}

We start by introducing a covariate-adaptive randomization procedure with $n$ units. Suppose that treatments $a \in \mathcal{A} = \{1,2,\dots,A\}$. Let $a=0$ be the control group and $\mathcal{A}_0 = \mathcal{A}\cup\{0\}$. Let $A_i$ be the indicator variable, such that $A_i = a$ indicates that the $i$th unit is assigned to the $a$th treatment. Let $n_a = \sum_{i=1}^n I(A_i=a)$ 
be the number of units in each treatment group. Let $B_i$ denote the stratum label, which takes values in $\mathcal{K} = \{1, 2, \dots, K\}$. Here, $K$ represents the total number of strata, which is fixed and finite. Define $X_i = (X_{i1},\dots,X_{ip})^{\rm T}$ as the $p$-dimensional additional baseline covariates not used in the randomization procedure. We consider a high-dimensional setting where $p$ tends to infinity as $n$ goes
to infinity. Let $p_{[k]} = \mathrm{pr}(B_i=k) >0$ be the expected proportion of stratum $k$ and $\pi_{[k]a} = \mathrm{pr}(A_i=a\mid B_i=k)$ be the expected proportion of treatment $a$ in stratum $k$. Let $n_{[k]} = \sum_{i=1}^n I(B_i=k)$ be the number of units in stratum $k$ and $n_{[k]a} =\sum_{i=1}^n I(A_i = a,B_i=k)$ be the number of units in stratum $k$ and treatment group $a$. Let $p_{n[k]} = n_{[k]}/n$ be the estimated proportion of stratum $k$. 

We use the Neyman-Rubin model to define potential outcomes and treatment effects \citep{Neyman1990, Rubin1974}. Let \{$Y_i(a), a\in \mathcal{A}_0$\} be the potential outcomes and $Y_i = \sum_{a\in\mathcal{A}_0} I(A_i=a)Y_i(a)$ be the observed outcomes. Let $ \{W_i\}_{i=1}^n = \{Y_i(0),\dots,Y_i(A),B_i, X_i\}_{i=1}^n$ be independent and identically distributed (i.i.d.) samples from the population distribution $W=\{Y(0),\dots,Y(A),B,X\}$. Our goal is to estimate the treatment effect $\tau_a = E\{Y_i(a) - Y_i(0)\}$ for all $a \in \mathcal{A}$. 

We denote the set of random variables with at least one positive stratum-specific variance as $\mathcal{R}_2 = \{V:\ \max_{k \in \mathcal{K}} \var(V|B_i=k)>0\}$. Also, we assume that the stratum-specific covariance matrix 
\[\Sigma_{[k]XX} = E[\{X_i - E(X_i|B_i=k)\}\{X_i - E(X_i|B_i=k)\}^{\rm T}|B_i=k],\ k=1,2,\dots,K,\]
is strictly positive definite. The following assumptions are made for the data generation and covariate-adaptive randomization procedures. 

\begin{assump}\label{ap1}
$E\{Y^2_i(a)\} < \infty $ and $Y_i(a) \in \mathcal{R}_2$, for all $a\in \mathcal{A}_0$.  $\{X_i\}_{i=1}^n$ are uniformly bounded by a constant, which means that there exists a constant $M$ independent of $n$, such that $\max_{i=1,\dots,n; j=1,\dots,p} |X_{ij}|\leq M$.
\end{assump}

\begin{assump}\label{ap2}\ 
\begin{enumerate}
    \item Conditional on  $\{B_1,\dots,B_n\}$, $\{A_1,\dots,A_n\}$ is independent  of $\{Y_i(0),\dots,Y_i(A), X_i\}_{i=1}^n$.
    \item $n_{[k]a}/n_{[k]} \stackrel{P}\to \pi_{[k]a}$ as $n \to \infty$, for all $a \in \mathcal{A}_0$ and $k \in \mathcal{K}$.
\end{enumerate}

\end{assump}

The assumptions are similar to those proposed in \citet{gu2023regression}. The difference for Assumption \ref{ap1} is that $X_i$ is high-dimensional, so we assume that $X_i$ is uniformly bounded, as motivated by \citet{liu2022lasso}. This assumption may be stringent for $X_i$; however, it helps relax the requirements for the approximation error. Assumption \ref{ap2}.1 requires that given the strata, the treatment assignment is independent of the potential outcomes and the additional covariates. Assumption \ref{ap2}.2 is the same as Assumption 2.2(b) in \citet{Bugni2019}. Several well-known covariate-adaptive randomization procedures satisfy this assumption, including stratified block randomization \citep{zelen1974randomization}, and stratified biased-coin randomization \citep{Shao2010}. In particular, simple randomization can be considered a special case of covariate-adaptive randomization, in which case there is one stratum and the treatment assignments are i.i.d Bernoulli variables. Moreover, when $\pi_{[k]a}$ is identical across strata, Pocock and Simon's minimization \citep{Pocock1975} also satisfies this assumption \citep{hu2022multi}.

Next, we introduce some notations. Let $r_i(a)$, for $i = 1,2,\dots,n,\ a\in\mathcal{A}_0$, be a transformed outcome, such as potential outcomes $Y_i(a)$, covariates $X_i$ or their linear combinations. \del{Also, we denote $\tilde{r}_i(a) = r_i(a) - E\{r_i(a)|B_i\}$.} The sample means of the transformed outcomes are defined as $\bar{r}_a = (1/n_a)\ \sum_{i=1}^n I(A_i=a)r_i(a)$ and $\bar{r}_{[k]a} =  (1/n_{[k]a})\ \sum_{i=1}^n I(A_i=a,B_i=k) r_i(a)$. The population variance of a transformed outcome $r_i(a)$ is denoted as $\sigma^2_{r(a)}$.  We define $\sigma^2_{[k]r(a)} = \var \{r_i(a)\mid B_i=k\}$ as the population variance of a transformed outcome in stratum $k$. 

\section{Lasso-adjusted estimator under multiple treatments}

The stratum-specific estimator under multiple treatments has been discussed in \citet{gu2023regression}. This estimator can guarantee an efficiency gain among the estimators they considered, regardless of whether the target allocation ratios across strata are the same or different. However, the OLS estimator does not work well in high-dimensional settings due to overfitting. Therefore, selecting covariates or some forms of regularization should be considered, such as lasso \citep{tibshirani1996regression}. \citet{liu2022lasso} considered the stratum-specific lasso-adjusted treatment effect estimator and derived its asymptotic normality when there are two treatments. In this section, we generalize the stratum-specific lasso-adjusted treatment effect estimator to multiple treatments and derive its asymptotic normality.

We define the stratum-specific projection coefficient $\ProjBeta$ as 
\[\ProjBeta = \arg\min_{\beta\in\mathbb{R}^p} E([\{Y_i(a) -E(Y_i(a)|B_i=k)\} -\{X_i -E(X_i|B_i=k)\}^{\rm T}\beta]^2 |B_i=k).\]
The stratum-specific lasso-adjusted vectors can be calculated as 
 \[\hat\beta_{[k]\rm lasso}(a) = \arg\min_{\beta \in \mathbb{R}^p} \frac{1}{n_{[k]a}}\sum\limits_{i=1}^n I(A_i=a, B_i = k) \{Y_i - \bar Y_{[k]a} - (X_i - \bar X_{[k]a})^{\rm T}\beta\}^2 + \lambda_{[k]a}\norm{\beta}{1}.\]
Therefore, the treatment effect estimator for each treatment $a$ is 
\[\hat{\tau}_{\rm lasso,a} = \sum\limits_{k \in \mathcal{K}} p_{n[k]}\big[\{\bar{Y}_{[k]a} - (\bar{X}_{[k]a} - \bar{X}_{[k]})^{\mathrm T} \hat{\beta}_{[k]\rm lasso}(a)\} -  \{\bar{Y}_{[k]0} - (\bar{X}_{[k]0} - \bar{X}_{[k]})^{\mathrm T} \hat{\beta}_{[k]\rm lasso}(0)\}\big].\]
and the stratum-specific lasso-adjusted treatment effect estimator is 
\[\hat\tau_{\rm lasso} = (\hat\tau_{\rm lasso,1},\hat\tau_{\rm lasso,2},\dots,\hat\tau_{\rm lasso,A})^{\rm T}.\]

To investigate the performance of $\hat\tau_{\rm lasso}$, we define the transformed outcome $r_{i,\rm proj}(a) = Y_i(a) - X_i^{\rm T} \ProjBeta$, for $i\in [k], A_i=a$. Let $\beta_{j,[k]\rm{proj}}(a)$ be the $j$th element of $\ProjBeta$ and $S_{[k]} = \bigcup_{a\in\mathcal{A}}\{j\in\{1,\dots,p\}:\ \beta_{j,[k]\rm{proj}}(a)\neq 0\}$ be the union of the support of $\ProjBeta$ for all $a \in\mathcal{A}_0$. Also, let $s_{[k]} = |S_{[k]}|$ be the number of relevant covariates in stratum $k$. We make the following assumptions within each stratum to obtain the $l_1$ convergence rate of vectors $\HLassoBeta$. 
\begin{assump}\label{ap3}
    The minimum eigenvalue of the stratum-specific covariance matrix $\Sigma_{[k]XX}$ is $\Lambda_{\min,[k]}$ and $\Lambda_{\min,[k]} \geq \kappa_l >0$, where $\kappa_l$ is a constant independent of $n$.
\end{assump}

\begin{assump}\label{ap4}
    There exists constants $c_{\lambda[k]}>0$, $c_M>1$ and sequence $M_{n[k]}\rightarrow \infty$ with $M_{n[k]}s^2_{[k]} (\log p)^2/n \rightarrow 0, k= 1,2,\dots, K$, such that the tuning parameters $\lambda_{[k]a}$ belongs to the interval
    \[\left[4\left(\frac{c_{\lambda[k]} p_{[k]} M_{n[k]}}{\pi_{[k]a}}\right)^{1 / 2}\left(\frac{\log p}{n}\right)^{1 / 2}, 4 c_M\left(\frac{c_{\lambda[k]} p_{[k]} M_{n[k]}}{\pi_{[k]a}}\right)^{1 / 2}\left(\frac{\log p}{n}\right)^{1 / 2}\right].\]
\end{assump}
Assumptions \ref{ap3}–\ref{ap4} closely resemble Assumptions 4 and 6 in \citet{liu2022lasso}. Specifically, Assumption \ref{ap3} can be identified as a more stringent requirement compared to Assumption 4 in \citet{liu2022lasso}. This adjustment aims to compromise the discussion within the framework of transfer learning. 

\begin{prop}\label{prop1}
    Under Assumptions \ref{ap1}--\ref{ap4}, if $r_{i,\rm proj}(a) \in \mathcal{R}_2$, $E\{r^2_{i,\rm proj}(a)\} <\infty$ for $a \in \mathcal{A}_0$, then $\norm{\hat{\beta}_{[k]\rm lasso}(a) - \ProjBeta}{1} = O_p[s_{[k]}(M_{n[k]}(\log p)/n)^{1/2}]$, for $k\in\mathcal{K}$ and $a \in \mathcal{A}_0$. Moreover, 
    \[\sqrt{n}(\hat\tau_{\rm lasso} - \tau) \stackrel{d}\rightarrow \operatorname{N}(0,V_{\rm proj}),\]
    where $V_{\rm proj} = V_{\tilde r_{\rm proj}} + V_{\tilde X} + V_{H\tilde Y}$,
    \begin{align*}
        V_{H\tilde{Y}} &= \sum\limits_{k \in \mathcal{K}} p_{[k]} \left(m_{[k]}(a) - m_{[k]}(0): a \in \mathcal{A}\right) \times  \left(m_{[k]}(a') - m_{[k]}(0): a' \in \mathcal{A}\right)^{\mathrm T}, \\
    V_{\tilde{r}_{\rm proj}} & = \sum\limits_{k \in \mathcal{K}}\frac{p_{[k]}}{\pi_{[k]0}} \sigma^2_{[k]\tilde{r}_{\rm proj}(0)} 1_A1_A^{\mathrm T} + \diag\left\{\sum\limits_{k \in \mathcal{K}} \frac{p_{[k]}}{\pi_{[k]a}} \sigma^2_{[k]\tilde{r}_{\rm proj}(a)}: a\in \mathcal{A}\right\},  \\
    V_{\tilde{X}} &= \Big\{\sum\limits_{k \in \mathcal{K}} p_{[k]}\{\beta_{[k]\rm proj}(a) - \beta_{[k]\rm proj}(0)\}^{\mathrm T} \Sigma_{[k]XX} \{\beta_{[k]\rm proj}(a') - \beta_{[k]\rm proj}(0)\}: (a,a') \in \mathcal{A}\times \mathcal{A}\Big\}.
    \end{align*}
    Here $1_A$ is an $A$-dimensional vector with ones and $m_{[k]}(a) = E[Y_i(a)|B_i=k] - E[Y_i(a)]$. 
\end{prop}
\begin{remark}
     A general theory in \citet{liu2022lasso} can guarantee the asymptotic normality for regression-adjusted estimator as long as
    \[\sqrt{n} (\bar X_{[k]1} - \bar X_{[k]0})^{\rm T} \{\hat \beta_{[k]}(a) - \beta_{[k]}(a)\} = o_p(1),\quad for\quad a=0,1\]
    is satisfied for some vector $\beta_{[k]}(a)$ and its estimator $\hat\beta_{[k]}(a)$. Under this condition, a regression-adjusted treatment effect estimator is asymptotic normal with mean $\tau = \tau_1 - \tau_0$.    Therefore, the crucial step for proving the asymptotic normality is to derive the $l_1$ convergence rate of the lasso-adjusted vectors. We prove our result by generalizing the existing theory to settings with multiple treatments.
\end{remark}

\begin{remark}\label{rem1}
    The unadjusted estimator was studied in \citet{Bugni2019}, which is \[\hat\tau_{\rm ben,a} = \sum_{k\in\mathcal{K}} p_{n[k]} (\bar Y_{[k]a} - \bar Y_{[k]0})\ \mathrm{for}\ a \in \mathcal{A}\] and 
    \[\hat\tau_{\rm ben} = (\hat\tau_{\rm ben,1},\hat\tau_{\rm ben,2},\dots,\hat\tau_{\rm ben,A})^{\rm T}.\] To improve the efficiency, \citet{gu2023regression} proposed a stratum-specific treatment effect estimator using linear regression in each stratum and treatment. Our proposed lasso-adjusted estimator shares the same asymptotic covariance matrix as the stratum-specific estimator, which had been certified to gain efficiency compared with the unadjusted estimator in \citet{gu2023regression}. This estimator will be used as the benchmark estimator in the numerical studies in Sections 6 and 7.
\end{remark}

\section{Transfer learning for treatment effect estimation}

\subsection{Context and algorithm}
The validity of the stratum-specific estimator requires a sufficient number of units in each stratum, which may not work under high-dimensional settings, especially when there are many small strata. To obtain valid inference, we use transfer learning by leveraging source data in the estimation. \del{In the context of transfer learning, besides the observation $\{Y_i, X_i, A_i, B_i\}_{i=1}^n$ comes from the target data,} we also have external trial data as source data. The treatments and the range of stratum labels in the source data are denoted as $\source{\mathcal{A}}_0$ and $\source{ \mathcal{K}}$. Let $\{\source W_i \}_{i=1}^{\source n}= \{\source Y_i(0),\dots,\source Y_i(A),\source B_i,\source X_i\}_{i=1}^{\source n}$ be $\source n$ i.i.d. samples from the source data population $\source W = \{\source Y(0),\dots,\source Y(A),\source B,\source X\}$, which is independent of $W$ and $(A_1,\dots,A_n)$. Similarly, the notations for source data are superscripted by $(s)$, such as $\source{r}_i(a)$ to indicate the transformed outcomes in source data and $\source n_{[k]}$ indicates the number of units in stratum $k$ in source data. We make the following assumptions about the source data. 
\begin{assump}\label{ap5}\ 
    \begin{enumerate}
        \item $\source X_i$ is a $p$-dimensional vector. $\source{\mathcal{A}}_0 = \mathcal{A}_0$ and $\source{ \mathcal{K}}=\mathcal{K}$.
        \item Source data $\{\source W_i\}_{i=1}^{\source n}$ satisfy Assumption \ref{ap1}. In details,  $E\left\{\left(\source{Y_i}\right)^2(a)\right\} < \infty $ and $\source Y_i(a) \in \mathcal{R}_2$, for all $a\in \source{\mathcal{A}}_0$.  $\{\source X_i\}_{i=1}^n$ are uniformly bounded by a constant, which means that there exists a constant $M$ independent of $\source n$ such that $\max_{i=1,\dots,\source n; j=1,\dots,p} |\source X_{ij}|\leq M$.
        \item The randomization procedure satisfies Assumption \ref{ap2}.
        In details,  conditional on  $\{\source B_1,\dots,\source B_n\}$, $\{\source A_1,\dots,\source A_n\}$ is independent with $\{\source Y_i(0),\dots,\source Y_i(A), \source X_i\}_{i=1}^{\source n}$. 
        \item $\source n_{[k]a}/\source n_{[k]} \stackrel{P}\to \source \pi_{[k]a}$ as $\source n \to \infty$, for all $a \in \source {\mathcal{A}}_0$ and $k \in \source {\mathcal{K}}$.
    \end{enumerate}
\end{assump}

Assumption \ref{ap5}.1 requires that the treatments and the stratum labels are identical in both source and target data. When two clinical trials use the same treatments, it implies that they use similar drugs, biomarkers, or treatment procedures. Likewise, identical stratum labels indicate that both trials use the same covariates, such as gender or age groups, for the randomization process. Assumption \ref{ap5}.2 is the same as Assumption \ref{ap1}. The upper bound $M$ is uniform across both source and target data. The covariate-adaptive randomization procedure depends on Assumption \ref{ap5}.3 and \ref{ap5}.4. Here we allow the randomization procedures in the two clinical trials can be different, as $\source \pi_{[k]a}$ may not be equal to $\pi_{[k]a}$. 

Our transfer learning algorithm is motivated by \citet{Li2022Transfer}. We apply the Oracle Trans-Lasso algorithm to estimate $\ProjBeta$. As an overview, we first derive the stratum-specific lasso-adjusted vectors in the source data. Then we correct the bias using target data, with an $l_1$ penalization on the bias. We define the stratum-specific projection coefficient in the source data as $\ProjBetaSource$, which is calculated as 
\[\ProjBetaSource = \arg\min_{\beta\in\mathbb{R}^p} E\bigg([\{\source Y_i(a) -E(\source Y_i(a)|\source B_i=k)\} -\{\source X_i -E(\source X_i|\source B_i=k)\}^{\rm T}\beta]^2 |\source B_i=k\bigg).\]
Also, we define the bias from $\ProjBetaSource$ to $\ProjBeta$ as  
%1. Why we want to use Trans-Lasso
%2. Benchmark -- use source directly
%3. Oracle Trans-lasso
%4. Regardless of h, we have a robust estimation (debias?)
%5. Simulation: l2betanorm, MSE?
%w.r.t nT, nS, s[k], h[k]?
\[\delta_{[k]}(a)= \ProjBeta-\ProjBetaSource .\]

Then we formally present our transfer learning algorithm in Algorithm 1.

\begin{table}[htbp]
    \centering
    \resizebox{1\textwidth}{!}{
    \begin{tabular}{l}
    \hline
    \textbf{Algorithm 1}: Transfer learning algorithm for treatment effect estimation under covariate-adaptive\\ randomization \\
    \hline \\
    \quad 1. Perform Lasso regression on source data in each corresponding stratum $k$ and treatment $a$. \\
    The coefficient estimators are derived by \\
    \\
    $\ProjBetaSourceE = \arg\min\limits_{\beta \in \mathbb{R}^p} \frac{1}{\source{n}_{[k]a}}\sum\limits_{i=1}^{\source{n}} I(\source{A}_i=a, \source{B}_i = k) \{\source{Y}_i - \source{\bar Y}_{[k]a} - (\source{X}_i -  \source{\bar X}_{[k]a})^{\rm T}\beta\}^2 + \source {\lambda}_{[k]a}\norm{\beta}{1}.$ \\
    \\
    \quad 2. Perform Lasso regression on the bias $\delta$ using target data. \\ 
    \\
    $\hat\delta_{[k]}(a) = \arg\min\limits_{\delta \in \mathbb{R}^p}\frac{1}{n_{[k]a}} \sum\limits_{i=1}^n I(A_i=a,B_i=k)\{Y_i - \bar Y_{[k]a} - (X_i - \bar X_{[k]a})^{\rm T}(\ProjBetaSourceE + \delta)\}^2 + \lambda_{[k]a}\norm{\delta}{1}.$\\
    \\
    \quad 3. Drive the coefficient estimator $\hat\beta_{[k]\rm tl}(a) = \hat\delta_{[k]}(a) + \ProjBetaSourceE $.\\
    \\
    \quad 4. Impute the coefficient estimators into the treatment effect estimator for each $a\in\mathcal{A}$: \\
    \\
    $\hat\tau_{\rm tl,a} =  \sum\limits_{k \in \mathcal{K}} p_{n[k]}\big[\{\bar{Y}_{[k]a} - (\bar{X}_{[k]a} - \bar{X}_{[k]})^{\mathrm T} \hat{\beta}_{[k]\rm tl}(a)\} -  \{\bar{Y}_{[k]0} - (\bar{X}_{[k]0} - \bar{X}_{[k]})^{\mathrm T} \hat{\beta}_{[k]\rm tl}(0)\}\big],$\\
    \\
and the transfer learning estimator is \\
\hspace{0.45\textwidth}$\hat\tau_{\rm tl} = (\hat\tau_{\rm tl,1}, \hat\tau_{\rm tl,2},\dots,\hat\tau_{\rm tl,A})^{\rm T}.$ \\
\hline
    \end{tabular}
    }
\end{table}

\subsection{Theoretical results}

We consider  $\delta_{[k]}(a)$ by its $l_1$ norm such that \[\max_{k\in\mathcal{K},\ a\in\mathcal{A}_0}\norm{\delta_{[k]}(a)}{1}\leq h.\] 
Recall that the transfer learning estimator for each treatment $a$ is 
\[\hat\tau_{\rm tl,a} =  \sum\limits_{k \in \mathcal{K}} p_{n[k]}\big[\{\bar{Y}_{[k]a} - (\bar{X}_{[k]a} - \bar{X}_{[k]})^{\mathrm T} \hat{\beta}_{[k]\rm tl}(a)\} -  \{\bar{Y}_{[k]0} - (\bar{X}_{[k]0} - \bar{X}_{[k]})^{\mathrm T} \hat{\beta}_{[k]\rm tl}(0)\}\big],\]
and the transfer learning estimator is
\[\hat\tau_{\rm tl} = (\hat\tau_{\rm tl,1}, \hat\tau_{\rm tl,2},\dots,\hat\tau_{\rm tl,A})^{\rm T}.\]
To derive the asymptotic behavior of $\hat\tau_{\rm tl}$, we first obtain the $l_1$ convergence rate of $\hat{\beta}_{[k]\rm tl}(a)$. The following assumptions are made for source data to reach our results.

\begin{assump}\label{ap6}
    The minimum eigenvalue of stratum-specific covariance matrix $\source\Sigma_{[k]XX}$ is $\source{\Lambda}_{\min,[k]}$ and $\source\Lambda_{\min,[k]} \geq \kappa_l >0$.
\end{assump}

\begin{assump}\label{ap7}
    There exists constants $\source c_{\lambda[k]}>0$, $\source c_M>1$ and sequence $\source M_{n[k]}\rightarrow \infty$ with $\source M_{n[k]}s^2_{[k]} (\log p)^2/\source n \rightarrow 0, k= 1,2,\dots, K$, such that the tuning parameters $\source \lambda_{[k]a}$ belongs to the interval
    \[\left[4\left(\frac{\source c_{\lambda[k]} \source p_{[k]} \source M_{n[k]}}{\source\pi_{[k]a}}\right)^{1 / 2}\left(\frac{\log p}{\source n}\right)^{1 / 2}, 4 c_M\left(\frac{\source c_{\lambda[k]} \source p_{[k]} \source M_{n[k]}}{\source\pi_{[k]a}}\right)^{1 / 2}\left(\frac{\log p}{\source n}\right)^{1 / 2}\right].\]
\end{assump}

Assumptions \ref{ap6} and \ref{ap7} are the source data versions of Assumptions \ref{ap3} and \ref{ap4} respectively. Note that $\source M_{n[k]}$ is the same sequence as $M_{n[k]}$ but with $\source n_{[k]}$ elements. The following theorem shows the $l_1$ convergence rate of $\hat\beta_{[k]\rm tl}(a)$.

\begin{thm}\label{cor1}
    Under Assumptions \ref{ap1}--\ref{ap3}, and \ref{ap5}--\ref{ap7}, suppose $h < s_{[k]}\sqrt{\log p/n}$, $\source n \gg n$ and $h(\log p/\source n)^{1/2} = o(1)$, tuning parameter $\lambda_{[k]a}$ belongs to the interval 
     \[\left[4\left(\frac{c_{\lambda[k]} p_{[k]} M_{n[k]}}{\pi_{[k]a}}\right)^{1 / 2}\left(\frac{\log p}{n}\right)^{1 / 2}, 4 c_M\left(\frac{c_{\lambda[k]} p_{[k]} M_{n[k]}}{\pi_{[k]a}}\right)^{1 / 2}\left(\frac{\log p}{n}\right)^{1 / 2}\right].\]
     where $M_{n[k]}\rightarrow \infty$ with $M_{n[k]}/n \rightarrow 0$. When $r_{i,\rm proj}(a) \in \mathcal{R}_2$ and $E\{r_{i,\rm proj}(a)\}^2 < \infty$ for $a \in \mathcal{A}_0$, we have
    \[\norm{\hat\beta_{[k]\rm tl}(a) - \ProjBeta}{1} = O_p\left(s_{[k]}\sqrt{\frac{\source M_{n[k]}\log p}{n+\source n}} + h\right).\]
\end{thm}

Theorem \ref{cor1} establishes the $l_1$ convergence rate of $\hat\beta_{[k]\rm tl}(a)$ under mild regularity conditions for $k\in\mathcal{K}, a \in \mathcal{A}_0$. We highlight the advantages of our method over the lasso-adjusted estimator discussed in Proposition \ref{prop1}. Theorem \ref{cor1} shows that the convergence rate is nearly $O_p\left(s{[k]}\sqrt{\source M_{n[k]}\log p/(n + \source n)}\right)$, allowing for less sparsity in the target model compared to the lasso-adjusted estimator. The conditions for achieving this improvement are $\source n \gg n$ and $h < s_{[k]}\sqrt{\log p/n}$. These conditions can be satisfied when the source data has large amounts of units and is similar to the target data. Meanwhile, the convergence rate and the sparsity condition are similar to the Oracle Trans-Lasso estimator in \cite{Li2022Transfer} except for a sequence $\source M_{n[k]}$. This sequence plays as the cost to relax the conditions on the transformed outcome $r_{i,\rm proj}(a)$. $\source M_{n[k]}$ can grow very slowly such that $\source M_{n[k]}/\source n$ approaches zero when $\source n$ is large enough, for example, $\source M_{n[k]} = \log\log \source n$. Therefore, the main requirement on the tuning parameter $\source \lambda_{[k]a}$ is of the \del{order $O\{(\log p/\source n)^{1/2}\}$.}

Now, we can obtain the asymptotic behaviour of $\hat\tau_{\rm tl}$.

\begin{thm}\label{thm1}
     Under Assumptions \ref{ap1}--\ref{ap3}, and \ref{ap5}--\ref{ap7}, suppose $h < s_{[k]}\sqrt{\log p/n}$, $\source n \gg n$ and $h(\log p)^{1/2} = o(1)$, tuning parameter $\lambda_{[k]a}$ satisfies the condition in Theorem \ref{cor1}. When $r_{i,\rm proj}(a) \in \mathcal{R}_2$ and $E\{r_{i,\rm proj}(a)\}^2 < \infty$ for $a \in \mathcal{A}_0$, we have
    \[\sqrt{n}(\hat\tau_{\rm tl} - \tau) \stackrel{d}\rightarrow \operatorname{N}(0,V_{\rm proj}).\]
\end{thm}

\begin{remark}
    We compare the result in Theorem \ref{thm1} with the result in Proposition \ref{prop1} for lasso. The requirement on the sparsity $s_{[k]}$ is relaxed to $s_{[k]}\log p = o(\sqrt{\source n/\source M_{n[k]}})$, when $\source n \gg n$, $h<s_{[k]}\sqrt{\log p/n}$. This result allows fewer units in the target data, especially when many strata exist and a small number of units exist in each stratum. Also, in typical high-dimensional regression models, the valid inference requires $s_{[k]} \log p \ll \sqrt{n}$. In Theorem \ref{thm1}, $s_{[k]}$ can be greater than $\sqrt{n}$ and of order $o(\sqrt{\source n})$.
\end{remark}

\begin{remark}
    We discuss the conditions on $h$ in Theorem \ref{thm1}. The typical condition for the inference using the debiased lasso \citep{zhang2014confidence} is $s_{[k]}\log p = o(\sqrt{n})$, which leads to $h$ is of order $o(1/\sqrt{\log p})$. Meanwhile, the larger $s_{[k]}$ is, the weaker the conditions on $h$. The condition $h(\log p)^{1/2}=o(1)$ coincides with Theorem 4.1 in \citet{li2022GGM}, which was required for the inference of the transfer learning estimator for the generalized linear model.
\end{remark}

\section{Consistent variance estimator}\label{sec4}

 A crucial step in drawing valid inference is constructing a consistent asymptotic variance estimator. Specifically, we consider the difference in the potential outcomes for any two treatments $b$ and $c$. Let $e_{bc}$ be an $A$-dimensional vector, such that all elements are zero, except for the $b$th element, which is 1, and the $c$th element, which is -1. Let $\tau_{bc} = E\{Y(b) - Y(c)\} = \tau_b - \tau_c,\ b,c\in \mathcal{A}$, be the real difference in potential outcomes, and $\hat{\tau}_{\dag,bc}$ be the estimator for $\tau_{bc}$ generated from the estimators above, for $\dag = \rm lasso, \rm tl$. Based on simple calculations,
\[\sqrt{n}(\hat{\tau}_{\dag,bc} - \tau_{bc})\stackrel{d}\rightarrow \operatorname{N}(0,\sigma^2_{\rm proj, bc}),\quad \dag = \rm lasso,\rm tl ,\]
where
\[\sigma_{\rm proj,bc}^2 = e^{\mathrm T}_{bc}V_{\rm proj}e_{bc} = V_{\rm proj}(b,b) - 2V_{\rm proj}(b,c)+V_{\rm proj}(c,c).\]
Here $V_{\rm proj}(i,j)$ denotes the element in the $i$th row and $j$th column of $V_{\rm proj}$. By using $r_i(a)$ as the transformed outcomes again and $\hat r_i(a)$ as an estimator of $r_i(a)$, $\hat r_i = \sum_{a\in\mathcal{A}_0}I(A_i=a) \hat r_i(a) $. Let $\bar{\hat r}_{[k]a} = 1/n_{[k]a} \sum_{i=1}^n I(A_i=a,B_i=k)\hat r_i$. For the estimators above, \del{we use $\hat r_{i,\dag} = Y_i - X_i^{\rm T}\hat \beta_{[k],\dag}(a)$, where $\dag = \rm lasso,\rm tl$, respectively, for $i \in [k],\ A_i=a$. We define
\[\hat{\sigma}^2_{r_{\dag}(a)} =  \sum\limits_{k \in \mathcal{K}} \frac{p_{n[k]}n_{[k]}}{n_{[k]a}} \bigg\{\frac{1}{n_{[k]a}}\sum\limits_{i=1}^n I(A_i=a,B_i=k)(\hat r_{i,\dag} - \bar{\hat r}_{[k]a,\dag})^2\bigg\},\quad \dag = \rm lasso,\rm tl,\]}
and
\[\hat{\sigma}^2_{bc,HY} = \sum\limits_{k \in \mathcal{K}} p_{n[k]} \bigg\{\bigg(\bar{Y}_{[k]b} - \sum\limits_{k' \in \mathcal{K}}p_{n[k']}\bar{Y}_{[k']b}\bigg)-\bigg(\bar{Y}_{[k]c}-\sum\limits_{k' \in \mathcal{K}}p_{n[k']}\bar{Y}_{[k']c}\bigg)\bigg\}^2.\]
Let $S_{[k]XX} = 1/n_{[k]}\ \sum_{i=1}^n I(B_i=k) (X_i-\bar{X}_{[k]})(X_i-\bar{X}_{[k]})^{\rm T} $ denote the stratum-specific sample covariance matrix. Let $\hat\sigma^2_{\dag,bc}$ be the estimator of $\sigma^2_{\rm proj,bc}$ for $\dag = \rm lasso, \rm tl$ respectively.

\begin{thm}\label{thm3}\ 
\begin{enumerate}
    \item Under Assumptions \ref{ap1}--\ref{ap4}, the variance estimator 
    \begin{multline*}
        \hat{\sigma}^2_{\rm lasso,bc} = \hat{\sigma}^2_{r_{\rm lasso}(b)} + \hat{\sigma}^2_{r_{\rm lasso}(c)} + \hat{\sigma}^2_{bc,HY} \\ +
        \sum\limits_{k \in \mathcal{K}}p_{n[k]}\big\{\hat{\beta}_{[k]\rm lasso}(b)-\hat{\beta}_{[k]\rm lasso}(c)\big\}^{\mathrm T} S_{[k]XX}\big\{\hat{\beta}_{[k]\rm lasso}(b)-\hat{\beta}_{[k]\rm lasso}(c)\big\}
    \end{multline*}
    is a consistent estimator for $\sigma^2_{\rm proj,bc}$.

    \item Under conditions in Theorem \ref{thm1}, the variance estimator 
    \begin{multline*}
        \hat{\sigma}^2_{\rm tl,bc} = \hat{\sigma}^2_{r_{\rm tl}(b)} + \hat{\sigma}^2_{r_{\rm tl}(c)} + \hat{\sigma}^2_{bc,HY} \\ +
        \sum\limits_{k \in \mathcal{K}}p_{n[k]}\big\{\hat{\beta}_{[k]\rm tl}(b)-\hat{\beta}_{[k]\rm tl}(c)\big\}^{\mathrm T} S_{[k]XX}\big\{\hat{\beta}_{[k]\rm tl}(b)-\hat{\beta}_{[k]\rm tl}(c)\big\}
    \end{multline*}
    is a consistent estimator for $\sigma^2_{\rm proj,bc}$.
\end{enumerate}

\end{thm}

\begin{remark}\label{rem2}
  A concern with Theorem \ref{thm1} is that the condition $h < s_{[k]}\sqrt{\log p/n}$ may not hold.  
However,  valid inference for $\hat\tau_{\rm tl}$ can still be provided using a debiased variance estimator. Assume there exists some coefficient vector $\beta_{[k]}(a)$ such that
    \[\sqrt{n} (\bar X_{[k]1} - \bar X_{[k]0})^{\rm T} \{\hat \beta_{[k]\rm tl}(a) - \beta_{[k]}(a)\} = o_p(1).\] 
    Then under Assumptions \ref{ap1} and \ref{ap2}, we can show that $\sqrt{n}(\hat\tau_{\rm tl}-\tau) $ is asymptotically normal with a covariance matrix depending on $\beta_{[k]}(a)$. At the same time, we can provide a consistent variance estimator for this new asymptotic variance as we did in this section. The related theorems and proofs are provided in the Appendix. A direct result is that when Assumptions \ref{ap5}-\ref{ap7} hold, $\hat\beta_{[k]\rm tl}(a)$ can be replaced by $\ProjBetaSourceE$ to obtain consistent treatment effect estimator and variance estimator. However, the efficiency of this source-only estimator could be lower. In the next section, we present simulations to compare the performance of these estimators.
\end{remark}

\section{Simulation}\label{sec7}

In this section, we evaluate the performance of our proposed estimators compared to a benchmark estimator across various scenarios. The benchmark estimator is the unadjusted estimator in \citet{Bugni2019}, as discussed in Remark \ref{rem1}. We compared the lasso-adjusted treatment effect estimator $\hat\tau_{\rm lasso}$, the transfer learning treatment effect estimator $\hat\tau_{\rm tl}$, and the source-only treatment effect estimator $\hat\tau_{\rm so}$ as we mentioned in Remark \ref{rem2}. The source-only treatment effect estimator for each treatment $a$ is defined as
\[\hat\tau_{\rm so,a} =  \sum\limits_{k \in \mathcal{K}} p_{n[k]}\big[\{\bar{Y}_{[k]a} - (\bar{X}_{[k]a} - \bar{X}_{[k]})^{\mathrm T} \ProjBetaSourceE\} -  \{\bar{Y}_{[k]0} - (\bar{X}_{[k]0} - \bar{X}_{[k]})^{\mathrm T} \source{\hat\beta}_{[k]\rm proj}(0)\}\big]\]
and $\hat\tau_{\rm so}$ is given by
\[\hat\tau_{\rm so} = (\hat\tau_{\rm so,1},\hat\tau_{\rm so,2},\dots,\hat\tau_{\rm so,A})^{\rm T}.\]
$\hat\tau_{\rm so}$ can be seen as we transfer the coefficient estimator in source data without any bias correction. 

The following model is used to generate the potential outcomes for both source and target data:
\[Y_i(a) = \mu_a + g_a(X_i) + \epsilon_{a,i},\ \mathrm{for}\ a\in\mathcal{A}_0,\]
where $X_i$, $g_a(X_i)$ are specified below. We consider $|\mathcal{A}_0|=3$, that is, 2 treatment groups and a control group. In each model, $(X_i,\epsilon_{0,i},\epsilon_{1,i},\epsilon_{2,i})$ are i.i.d. $\epsilon_{a,i}$ follows standard normal distribution, for $a=0,1,2$. Covariates other than $X_i$ are generated such that the total dimensional is $p=100$ to consider the high-dimensional setting. The details of the models are as follows: \\

Model 1: We consider $s$-dimensional $X_i$ such that $g_a(X_i) = \beta_1 X_{i1}+ \sum_{j=2}^s \beta_j X_{i1}X_{ij}$, for $a = 0,1,2$, where $X_{i1}$ takes value in $\{1,2\}$ with probabilities 0.4 and 0.6, respectively. $X_{ij}\sim \mathrm{Unif}[-2,2]$, for $j=2,3,\dots,s$ and they are independent with each other. The coefficient $\beta$ will be specified later for source data and target data separately. Here $X_{i1}$ is used for randomization, resulting in two strata. The additional covariates are independent of $X_{i}$ and follow a multivariate normal distribution with mean zero and a diagonal covariance matrix where all diagonal elements are 2. \\

Model 2: Similar to Model 1, we consider $s$-dimensional $X_i$, where $X_{i1}$ takes value in $\{1,2\}$ with probabilities 0.4 and 0.6, respectively. $X_{ij}\sim \mathrm{Beta}(2,2)$ for $j=2,3,\dots,s$, and they are independent with each other. For different treatments, we set $g_0(X_i) = \beta_1 X_{i1}+ \sum_{j=2}^s \beta_j X_{i1}(X_{ij}-0.5)$, $g_1(X_i) = \beta_1 X_{i1}+ \sum_{j=2}^s \beta_j X_{i1}(X^2_{ij}-0.3)$, and $g_2(X_i)=g_0(X_i)$. The coefficients $\beta$ and the additional covariates are set the same as in Model 1.\\

The source and target data use either Model 1 or Model 2 for generating potential outcomes. The details for each case are as follows:

Case 1: Both the source and target data use Model 1. We examine the results as the bias increases in a linear model setting. Let $s = \lceil n^r \rceil$ where $r = 0,0.05,\dots,0.7$. In target data, $\beta_j = 2$ for $j=1,2,\dots,s$ and in source data, $\source \beta_j = 2+\frac{hj}{s}$ for $j=1,2,\dots,s$, $h=0.2,0.5,1$. This setup leads to $\norm{\beta-\source \beta}{1} = h(s+1)/2$.\\ 

Case 2: We use Model 2 as the source data and Model 1 as the target data. This case demonstrates the compatibility of our transfer learning estimator when the source data is misspecified. The coefficients $\beta$, $\source \beta$ and dimension $s$ are the same as in Case 1.\\ 

Case 3: We use Model 1 as the source data and Model 2 as the target data. This case illustrates the robustness of our proposed estimators when the target data is misspecified. The coefficients $\beta$, $\source \beta$ and dimension $s$ are the same as in Case 1.\\ 

\begin{figure}[htbp]
\begin{subfigure}{1\textwidth}
    \centering
    \caption{}\label{fig1a}
    \includegraphics[width=0.87\linewidth,height=0.38\linewidth]{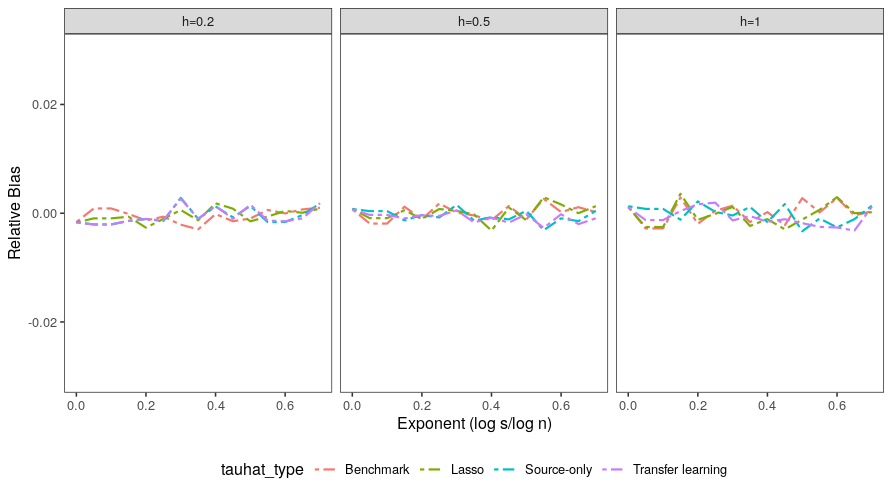}
 
\end{subfigure}

\begin{subfigure}{1\textwidth}
    \centering
    \caption{}\label{fig1b}
    \includegraphics[width=0.85\linewidth,height=0.38\linewidth]{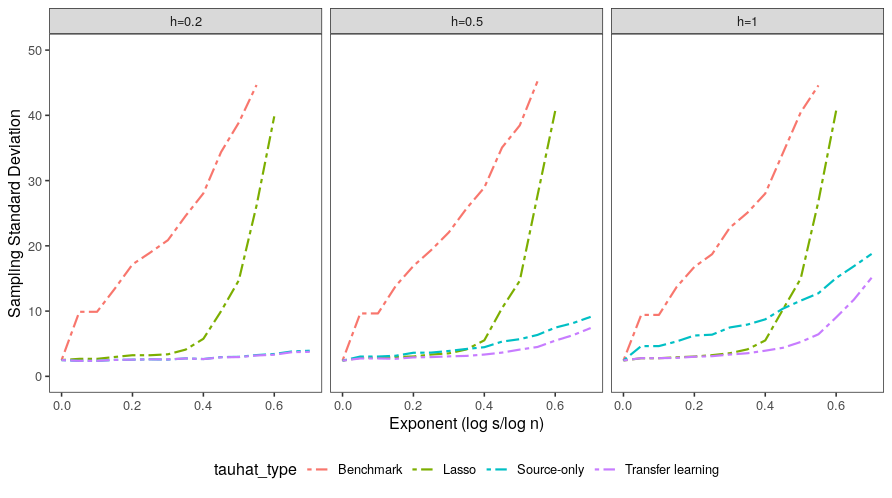}
\end{subfigure}
\begin{subfigure}{1\textwidth}
    \centering
    \caption{}\label{fig1c}
    \includegraphics[width=0.85\linewidth,height=0.38\linewidth]{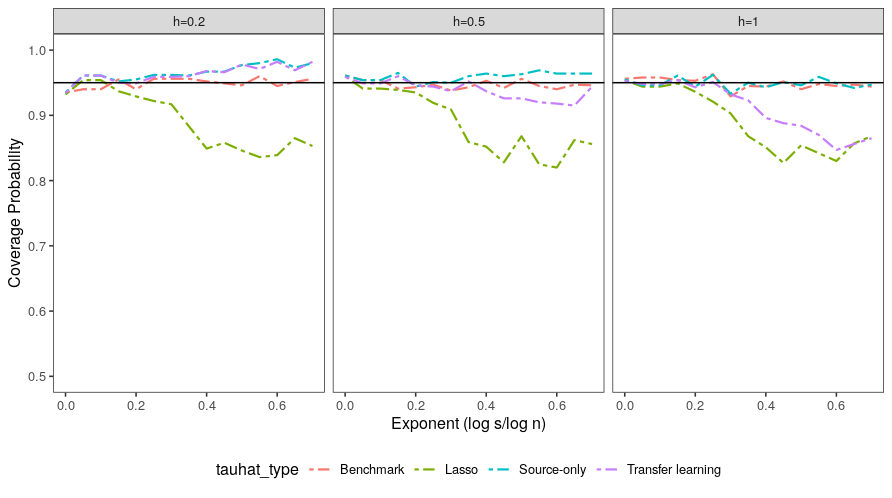}
\end{subfigure}
\caption{Simulations in Case 1 under stratified block randomization. Each column represents different $h$ in the simulation settings, resulting in different $l_1$ norms of the bias. (a) Relative bias for $\hat\tau_{\rm ben}$, $\hat\tau_{\rm lasso}$, $\hat\tau_{\rm so}$ and $\hat\tau_{\rm tl}$. (b) Sampling standard deviation for $\hat\tau_{\rm ben}$, $\hat\tau_{\rm lasso}$, $\hat\tau_{\rm so}$ and $\hat\tau_{\rm tl}$. (c) Coverage probability for $\hat\tau_{\rm ben}$, $\hat\tau_{\rm lasso}$, $\hat\tau_{\rm so}$ and $\hat\tau_{\rm tl}$.}\label{fig1}
 \end{figure}

\begin{figure}[htbp]
\begin{subfigure}{1\textwidth}
    \centering
        \caption{}
    \includegraphics[width=0.87\linewidth,height=0.38\linewidth]{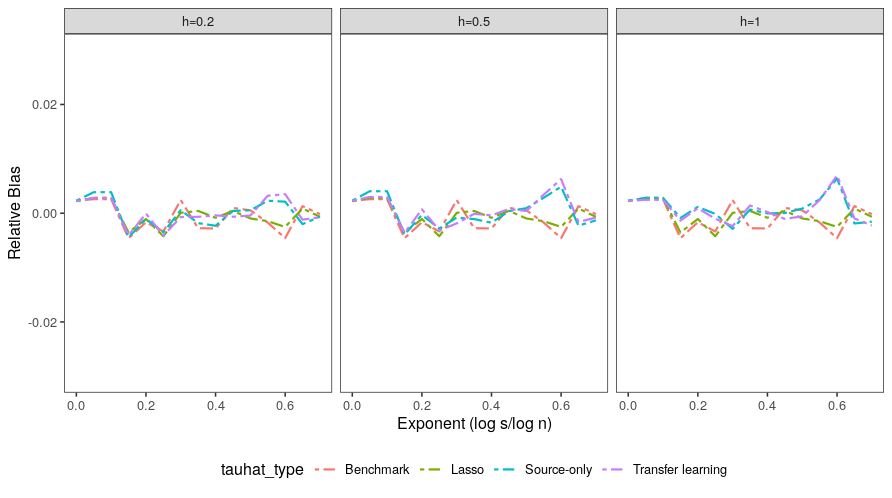}

\end{subfigure}

\begin{subfigure}{1\textwidth}
    \centering
        \caption{}
    \includegraphics[width=0.85\linewidth,height=0.38\linewidth]{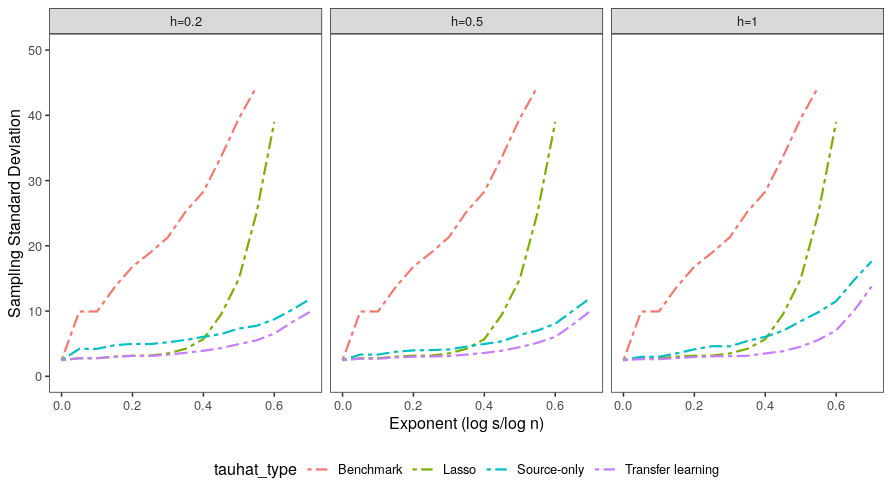}

\end{subfigure}
\begin{subfigure}{1\textwidth}
    \centering
        \caption{}
    \includegraphics[width=0.85\linewidth,height=0.38\linewidth]{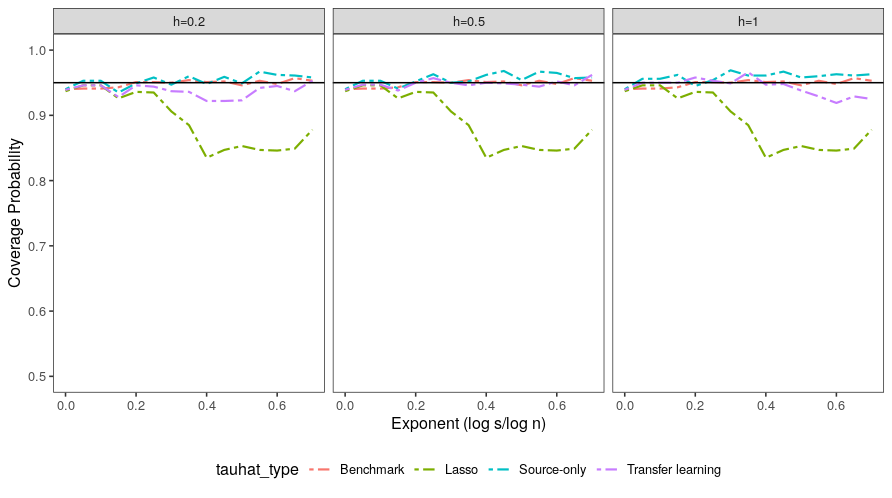}

\end{subfigure}
\caption{Simulations in Case 2 under stratified block randomization. Each column represents different $h$ in the simulation settings, resulting in different $l_1$ norms of the bias. (a) Relative bias for $\hat\tau_{\rm ben}$, $\hat\tau_{\rm lasso}$, $\hat\tau_{\rm so}$ and $\hat\tau_{\rm tl}$. (b) Sampling standard deviation for $\hat\tau_{\rm ben}$, $\hat\tau_{\rm lasso}$, $\hat\tau_{\rm so}$ and $\hat\tau_{\rm tl}$. (c) Coverage probability for $\hat\tau_{\rm ben}$, $\hat\tau_{\rm lasso}$, $\hat\tau_{\rm so}$ and $\hat\tau_{\rm tl}$.}\label{fig2}
 \end{figure}

\begin{figure}[htbp]
\begin{subfigure}{1\textwidth}
    \centering
        \caption{}
    \includegraphics[width=0.87\linewidth,height=0.38\linewidth]{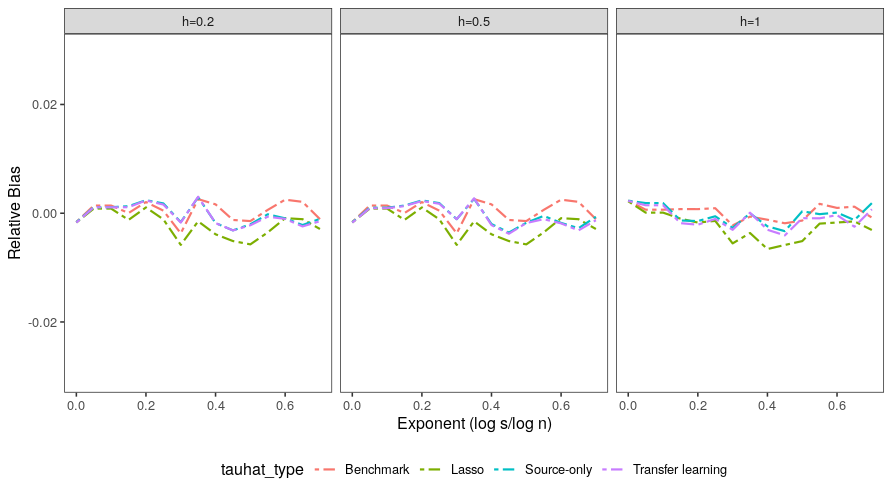}

\end{subfigure}

\begin{subfigure}{1\textwidth}
    \centering
        \caption{}
    \includegraphics[width=0.85\linewidth,height=0.38\linewidth]{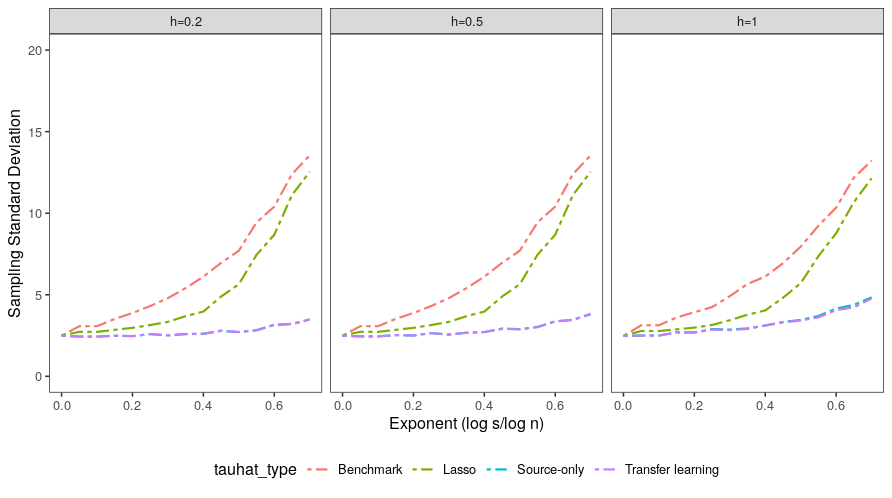}

\end{subfigure}
\begin{subfigure}{1\textwidth}
    \centering
        \caption{}
    \includegraphics[width=0.85\linewidth,height=0.38\linewidth]{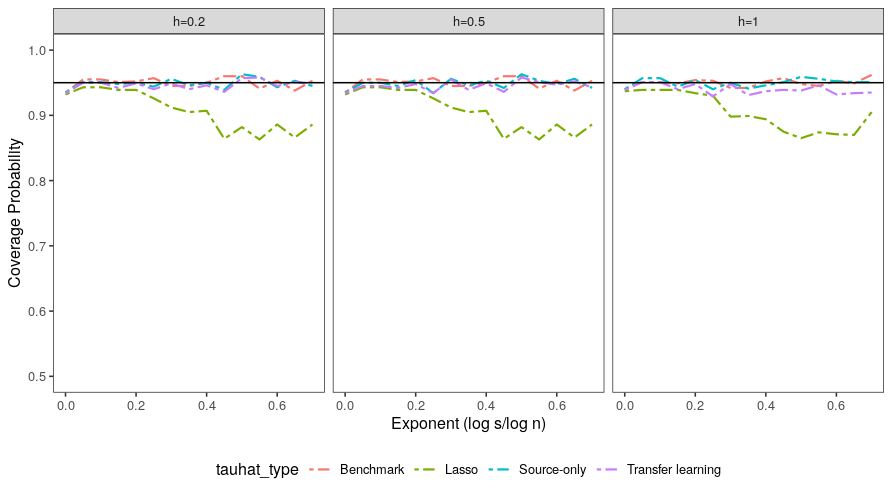}

\end{subfigure}
\caption{Simulations in Case 3 under stratified block randomization. Each column represents different $h$ in the simulation settings, resulting in different $l_1$ norms of the bias. (a) Relative bias for $\hat\tau_{\rm ben}$, $\hat\tau_{\rm lasso}$, $\hat\tau_{\rm so}$ and $\hat\tau_{\rm tl}$. (b) Sampling standard deviation for $\hat\tau_{\rm ben}$, $\hat\tau_{\rm lasso}$, $\hat\tau_{\rm so}$ and $\hat\tau_{\rm tl}$. (c) Coverage probability for $\hat\tau_{\rm ben}$, $\hat\tau_{\rm lasso}$, $\hat\tau_{\rm so}$ and $\hat\tau_{\rm tl}$.}\label{fig3}
 \end{figure}

We represent the simulation results for four treatment effect estimators under stratified block randomization. The target data size is $n = 300$ and the source data size is $\source n = 1200$. The allocation ratio is 1:1:1 among treatments, with a block size of 6. Our estimators are calculated in $R=2000$ replicates. The performances are measured by relative bias, standard deviation, and coverage probability of 95\% confidence interval. The relative bias is defined as $\sqrt{n} R^{-1} \sum_{r=1}^R (\hat\tau_{\dag,r} - \tau)/\hat\sigma$, where $\hat\tau_{\dag,r}$ denotes any of our treatment effect estimators in the $r$th replicate, $\tau$ is the true value of treatment effect, and $\hat\sigma$ is the sampling standard deviation. An additional result for unequal allocation ratio, which is 1:1:1 in the first strata and 2:2:1 in another strata, is in the Appendix. The results are summarized as follows:

\begin{itemize}
    \item For all cases, the relative bias for all estimators is close to 0.
    \item In Figure \ref{fig1b}, the sampling standard deviation (SD) for all estimators increases when $s$ increases. When $h$ increases, the difference in the sampling standard deviation between $\hat\tau_{\rm so}$ and $\hat\tau_{\rm lasso}$ increases as the source data becomes far away from the target data. The difference can be eased by our transfer learning approach. As we expected, when $s$ is small, the sampling standard deviation for $\hat\tau_{\rm tl}$ is similar to $\hat\tau_{\rm lasso}$, which verifies Theorem \ref{thm1}; when $s$ is sufficient large, $\hat\tau_{\rm tl}$ shows an efficiency gain compared to $\hat\tau_{\rm lasso}$.
    \item In Figure \ref{fig1c}, $\hat\tau_{\rm lasso}$ shows an incredible decrease in coverage probability when $s$ increases, which may be caused by the overfitting when the condition ($s\log p < \sqrt{n}$) is violated, leading to an inconsistent variance estimator. As we have mentioned in Remark \ref{rem2}, the coverage probability is approximately  95\% for $\hat\tau_{\rm so}$. For $\hat\tau_{\rm tl}$, the estimation is conservative when $h=0.2$. When $h=0.5$ and $s$ increases, $\hat\tau_{\rm tl}$ maintains a 95\% coverage probability as the conditions in Theorem \ref{thm1} hold. Moreover, even when the condition is violated at $h=1$, $\hat\tau_{\rm tl}$ shows a slower decrease in coverage probability compared to $\hat\tau_{\rm lasso}$. 
    \item In Figures \ref{fig2} and \ref{fig3}, the underlying model is misspecified from the linear model for source data or target data, respectively. The results for all measures are consistent with those presented in Figure \ref{fig1}, which shows that our transfer learning algorithm is robust regardless of whether the target data or the source data is misspecified. Especially our transfer learning approach performs better in Cases 2 and 3, with the coverage probabilities around 95\%. This confirms that our proposed approach can provide valid inference when conditions are satisfied.
\end{itemize}

\section{Clinical trial example}

In this section, we consider an example of a clinical trial solely to generate synthetic data to illustrate the capability of our proposed estimators to improve efficiency compared with the benchmark and provide valid inference. We use synthetic data from a trial using nefazodone and the cognitive behavioral-analysis system of psychotherapy (CBASP) \citep{keller2000comparison}. The trial aimed to compare the efficacy of nefazodone, CBASP, and their combination in the treatment of chronic depression. We consider the combination as the control group indexed as 0, nefazodone as the treatment indexed as 1, and the CBASP as the treatment indexed as 2. \del{The outcome of interest was FinalHAMD, the final score of the 24-item Hamilton rating scale for depression.} To generate the synthetic data, we first fit the data with a nonparametric spline using the function \texttt{bigssa} from the \texttt{R} package \texttt{bigspline}. Similar to \citet{liu2022lasso}, we stratify the data using GENDER and select eight covariates to fit the model: AGE, HAMA SOMATI, HAMD17, HAMD24, HAMD COGIN, Mstatus2, NDE and TreatPD, which are detailed in Table \ref{tab1}. We take linear, quadratic, cubic, and interaction terms for continuous covariates, linear and interaction terms for binary covariates, and interaction terms for the above two sets of coordinates as the covariates resulting in $p = 135$ for treatment effect estimators other than the benchmark estimator.

\begin{figure}[htbp]
    \centering
    \includegraphics[scale=0.65]{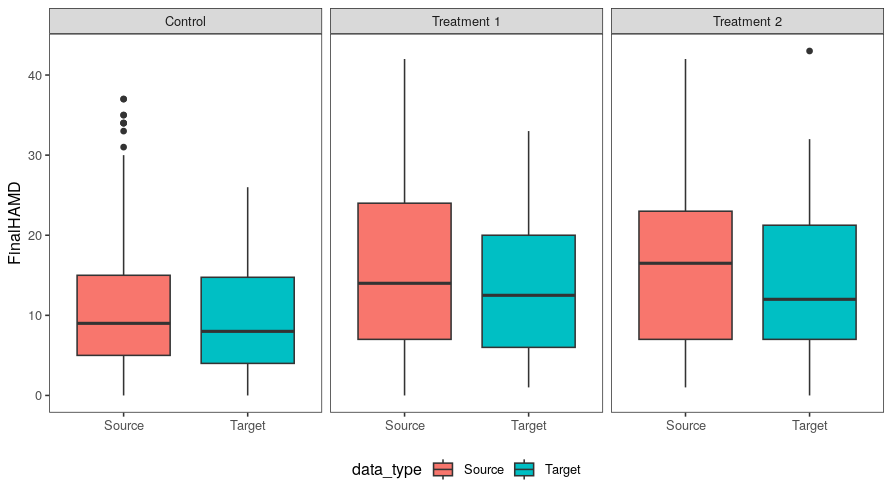}
    \caption{Difference in \del{potential outcomes (FinalHAMD)} between source data and target data}
    \label{fig4}
\end{figure}

To generate the source and target data, we divide the data by HAMD17 such that the target population has HAMD17 less than 18 and the source population has HAMD17 greater than or equal to 18. The purpose is to create a difference in outcomes between the source and target data, as shown by a boxplot in Figure \ref{fig4}. Then, we sample 300 units among the target population with replacement as the target dataset. Similarly,  we sample 1200 units from the source population with replacement as the source data. Next, we implement stratified block randomization on both the source and target data, using GENDER as the stratified variable with an allocation ratio of 1:1:1 for treatments and a block size of 6. We compare the performance of four estimators in Section \ref{sec7}. The treatment effects and variances are evaluated based on 1000 replicates. The coverage probabilities of the 95\% confidence interval are assessed based on the estimates and variance estimators. The results are shown in Table \ref{tab2}.

As shown in Table \ref{tab2}, the treatment effect estimation for all four estimators is similar with positive results. The source-only estimator $\hat\tau_{\rm so}$ has the worst performance in terms of efficiency, which may decrease the efficiency compared with the benchmark estimator $\hat\tau_{\rm ben}$ (variance reduction: -35.08\%). Both $\hat\tau_{\rm lasso}$ and $\hat\tau_{\rm tl}$ show efficiency gains compared with $\hat\tau_{\rm ben}$ as the variance reduction range from 25.53\% to 40.19\%, and $\hat\tau_{\rm tl}$ also shows an efficiency gain compared with $\hat\tau_{\rm lasso}$. Moreover, $\hat\tau_{\rm tl}$ and $\hat\tau_{\rm so}$ have better performance on coverage probabilities than $\hat\tau_{\rm lasso}$, which are closer to 95\%.   

\begin{table}[htbp]
    \caption{Description of baseline covariates}\label{tab1}
    \centering
    \begin{tabular}{ll}
    \hline
Variable & Description\\
\hline
AGE &Age of patients in years \\
GENDER & 1 female and 0 male\\
HAMA\_SOMATI & HAMA somatic anxiety score\\
HAMD17 & Total HAMD-17 score\\
HAMD24 & Total HAMD-24 score\\
HAMD\_COGIN & HAMD cognitive disturbance score\\
Mstatus2 & Marriage status: 1 if married or living with someone and\\
& 0 otherwise\\
NDE & Number of depressive episodes \\
TreatPD & Treated past depression: 1 yes and 0 no\\
\hline
\multicolumn{2}{l}{Note: HAMD, Hamilton Depression Scale; HAMA, Hamilton Anxiety Scale.} 
    \end{tabular}

\end{table}

\begin{table}[htbp]
\caption{Estimates, sampling variances, variance reductions and coverage probabilities under stratified block randomization for synthetic Nefazodone CBASP trial data.}\label{tab2}
\resizebox{1\textwidth}{!}{
\begin{tabular}{lllllllll}
\hline
                          & \multicolumn{4}{c}{$\tau_{1}$}           & \multicolumn{4}{c}{$\tau_{2}$}           \\ \hline
Estimator                 & Estimate & Variance ($\times 10^{-2}$)           & VR (\%) & CP    & Estimate & Variance ($\times 10^{-2}$)          & VR (\%) & CP    \\ \hline
$\hat\tau_{\rm ben}$ & 4.46     & 15.97
 & ---      & 0.956 & 4.29     & 22.7
 & ---      & 0.954 \\
 $\hat\tau_{\rm so}$   & 4.44     & 21.57
 & -35.08  & 0.950 & 4.30     & 19.13
 & 15.75   & 0.954 \\ 
$\hat\tau_{\rm lasso}$    & 4.46     & 11.89
 & 25.53   & 0.933 & 4.30     & 15.38
 & 32.25   & 0.913 \\
$\hat\tau_{\rm tl}$       & 4.46     & 11.28
 & 29.36   & 0.944 & 4.29     & 13.58
 & 40.19   & 0.950 \\ \hline
\multicolumn{9}{l}{Note: CI, confidence interval; VR, variance reduction; CP, coverage probability} 
\end{tabular}
}

\end{table}

\section{Discussion}

In this paper, we incorporate external trial data into the inference under covariate-adaptive randomization by transfer learning. We first generalize the lasso-adjusted stratum-specific treatment effect estimator under multiple treatments. \fix{Then, we apply a transfer learning approach to develop the transfer learning estimator. The theoretical results show that our proposed estimator can enhance the lasso-adjusted stratum-specific estimator to have faster $l_1$ convergence rate of regression coefficient estimators and weaken the requirements on the sample size for asymptotic normality, when the source data and target data are similar under mild conditions.} We also proposed a consistent variance estimator to provide valid inference. The simulation results and a clinical trial example confirm our findings. By proposing an effective transfer algorithm, offering theoretical guarantees, and developing a robust inference method, our study sheds light on the potential of transfer learning in analyzing randomized clinical trials.

% Future work

Our transfer learning approach requires that the source data and target data are similar concerning the $l_1$ norm of the difference in regression coefficients. In practice, the external trial data may substantially differ from the current trial data. Consequently, selecting units from the external trial data that align with the current trial is \fix{crucial but may complicate inference} due to the correlated selection procedure across both datasets. Therefore, identifying similar parts is a potential future work for our research. Meanwhile, the efficiency gain of the treatment effect estimation can be improved by estimating the underlying model between the response and the baseline covariates \citep{tu2023unified}. A potential improvement for our work is to replace the linear model with a general function for covariates. In this setting, future works may focus on characterizing the difference between the source data and the target data and proposing a proper transfer learning approach \fix{to develop a robust treatment effect estimator and ensure valid inference}. Moreover, potential works can focus on adapting transfer learning to estimate other types of treatment effects, such as quantile treatment effects and interaction effects.

% Why different
\bibstyle{apalike}
\bibliography{reference}

\renewcommand\thefigure{A\arabic{figure}} 
\newpage
\section*{Appendix}
\setcounter{figure}{0}    
\appendix

\newtheorem{appAssump}{Assumption}[section]

\newtheorem{appRemark}{Remark}[section]

\newtheorem{appProp}{Proposition}[section]

\section{General theory for regression adjustment under covariate-adaptive randomization with multiple treatments}

We develop a general theory for regression-adjusted treatment effect estimators, which we can apply to the lasso-adjusted and transfer learning estimators. 

\begin{appAssump}\label{ap}
     For $k=1,2,\dots,K$ and $a \in \mathcal{A}_0$, there exists coefficient vectors $\beta_{[k]}(a)$ such that
    \begin{align*}
        \sqrt{n} (\bar X_{[k]a} - \bar X_{[k]})^{\rm T}\{\hat \beta_{[k]}(a) - \beta_{[k]}(a)\} = o_p(1).
    \end{align*}
\end{appAssump}

\begin{appRemark}
    We develop the asymptotic normality by first deriving the asymptotic normality for the potential outcomes $\tau^* = (\tau^*_0,\dots,\tau^*_A)^{\rm T}$ and then using linear transformation to obtain the result, where $\tau^*_a = E\{Y_i(a)\}$ for $a \in\mathcal{A}_0$. Therefore, Assumption \ref{ap} is the remaining term for the first step and needs to be $o_p(1)$.
\end{appRemark}

\begin{appRemark}
    When the dimension $p$ is fixed, $\sqrt{n}(\bar X_{[k]a} - \bar X_{[k]}) = O_p(1)$ by \citet{Bugni2018}. Therefore, when $\hat \beta_{[k]}(a)$ element-wisely converges to $\beta_{[k]}(a)$, Assumption \ref{ap} holds. Remark 2 in \citet{gu2023regression} also explained this result when using the stratum-common estimator. However, the OLS estimator may not satisfy this condition when $p$ is diverging or even larger than the sample size.  
\end{appRemark}

Let $\hat \tau_{\rm{gen}}$ be a general estimator for treatment effect $\tau$.
\[\hat{\tau}_{\rm{gen},a} = \sum\limits_{k \in \mathcal{K}} p_{n[k]}\big[\{\bar{Y}_{[k]a} - (\bar{X}_{[k]a} - \bar{X}_{[k]})^{\mathrm T} \hat{\beta}_{[k]}(a)\} -  \{\bar{Y}_{[k]0} - (\bar{X}_{[k]0} - \bar{X}_{[k]})^{\mathrm T} \hat{\beta}_{[k]}(0)\}\big]\]
and 
\[\hat\tau_{\rm{gen}} = (\hat\tau_{\rm{gen},1},\dots,\hat\tau_{\rm{gen},A}).\]
Also, we define the transformed outcome $r_{i,\rm{gen}}(a) = Y_i(a) - X^{\rm T}_i\beta_{[k]}(a)$, for $i\in [k], A_i=a$.
\begin{appProp}\label{prop2}
    Under Assumptions \ref{ap1},\ref{ap2} and \ref{ap}, we have
    \[\sqrt{n} (\hat\tau_{\rm{gen}} -\tau) \stackrel{d}\rightarrow \operatorname{N}(0,V_{\rm{gen}}),\]
    where $V_{\rm{gen}} = V_{H\tilde Y}+V_{\tilde{r_{\rm{gen}}}} + V_{\tilde X} + V^{\rm T}_{\tilde{r}_{\rm{gen}}\tilde{X}}$, and 
\begin{align*}
    V_{H\tilde{Y}} &= \sum\limits_{k \in \mathcal{K}} p_{[k]} \left(m_{[k]}(a) - m_{[k]}(0): a \in \mathcal{A}\right) \times  \left(m_{[k]}(a') - m_{[k]}(0): a' \in \mathcal{A}\right)^{\mathrm T}, \\
    V_{\tilde{r}_{\rm{gen}}} & = \sum\limits_{k \in \mathcal{K}}\frac{p_{[k]}}{\pi_{[k]0}} \sigma^2_{[k]\tilde{r}_{\rm{gen}}(0)} 1_A1_A^{\mathrm T} + \diag\left\{\sum\limits_{k \in \mathcal{K}} \frac{p_{[k]}}{\pi_{[k]a}} \sigma^2_{[k]\tilde{r}_{\rm{gen}}(a)}: a\in \mathcal{A}\right\},  \\
    V_{\tilde{X}} &= \Big\{\sum\limits_{k \in \mathcal{K}} p_{[k]}\{\beta_{[k]}(a) - \beta_{[k]}(0)\}^{\mathrm T} \Sigma_{[k]XX} \{\beta_{[k]}(a') - \beta_{[k]}(0)\}: (a,a') \in \mathcal{A}\times \mathcal{A}\Big\},\\
    V_{\tilde{r}_{\rm{gen}}\tilde{X}} & = \Big\{\sum\limits_{k\in\mathcal{K}}p_{[k]}\{\beta_{[k]}(a')-\beta_{[k]}(0)\}^{\mathrm T}\big[(\Sigma_{[k]XY(a)}-\Sigma_{[k]XY(0)})- \Sigma_{[k]XX}\{\beta_{[k]}(a)-\beta_{[k]}(0)\}\big]: \\
    &(a,a')\in \mathcal{A}\times\mathcal{A}\Big\}.
\end{align*} 
Here $1_A$ is an $A$-dimensional vector with ones and $m_{[k]}(a) = E[Y_i(a)|B_i=k] - E[Y_i(a)]$. 
\end{appProp}

\begin{proof}
    The asymptotic normality can be derived from Theorem 2 in \citet{gu2023regression} under Assumption \ref{ap} with minor changes. 
    Recall that 
    \[\hat{\tau}_{\rm{gen},a} = \sum\limits_{k \in \mathcal{K}} p_{n[k]}\big[\{\bar{Y}_{[k]a} - (\bar{X}_{[k]a} - \bar{X}_{[k]})^{\mathrm T} \hat{\beta}_{[k]}(a)\} -  \{\bar{Y}_{[k]0} - (\bar{X}_{[k]0} - \bar{X}_{[k]})^{\mathrm T} \hat{\beta}_{[k]}(0)\}\big].\]
    Then we have
    \begin{align*}
        \hat{\tau}_{\rm{gen}} &= \Big(\sum\limits_{k \in \mathcal{K}} p_{n[k]}\big[\{\bar{Y}_{[k]a} - (\bar{X}_{[k]a} - \bar{X}_{[k]})^{\rm T} \hat\beta_{[k]}(a)\} -  \{\bar{Y}_{[k]0} - (\bar{X}_{[k]0} - \bar{X}_{[k]})^{\rm T} \hat\beta_{[k]}(0)\}\big]: a \in \mathcal{A}\Big) \\
&= \Big(\sum\limits_{k \in \mathcal{K}} p_{n[k]}\big[\{\bar{Y}_{[k]a} - (\bar{X}_{[k]a} -E(X_i|B_i=k))^{\rm T} \beta_{[k]}(a)\} \\ 
&-  \{\bar{Y}_{[k]0} - (\bar{X}_{[k]0} - E(X_i|B_i=k))^{\rm T} \beta_{[k]}(0)\}\big]: a\in \mathcal{A}\Big) \\ 
&- \Big(\sum\limits_{k \in \mathcal{K}} p_{n[k]} \{E(X_i|B_i=k) - \bar{X}_{[k]}\}^{\rm T}\{\beta_{[k]}(a) - \beta_{[k]}(0)\}: a \in \mathcal{A}\Big) \\
&+ \Big(\sum\limits_{k \in \mathcal{K}} p_{n[k]}[(\bar{X}_{[k]a} - \bar{X}_{[k]})^{\rm T} \{\hat\beta_{[k]}(a) - \beta_{[k]}(a)\} - (\bar{X}_{[k]0} - \bar{X}_{[k]})^{\rm T} \{\hat\beta_{[k]}(0) - \beta_{[k]}(0)\}]: a\in \mathcal{A}\Big). \\
    \end{align*}    
    The proof falls to show that the first two terms are asymptotic normal and the last term is $o_p(n^{-1/2})$. By Assumption \ref{ap4}, we have 
    \[(\bar X_{[k]a} - \bar X_{[k]})^{\rm T}\{\hat \beta_{[k]}(a) - \beta_{[k]}(a)\} = o_p(n^{-1/2})\]
    for all $a \in \mathcal{A}_0$. As the stratum size $K$ is finite, we have
    \[\sum\limits_{k \in \mathcal{K}} p_{n[k]}[(\bar{X}_{[k]a} - \bar{X}_{[k]})^{\rm T} \{\hat\beta_{[k]}(a) - \beta_{[k]}(a)\} - (\bar{X}_{[k]0} - \bar{X}_{[k]})^{\rm T} \{\hat\beta_{[k]}(0) - \beta_{[k]}(0)\}] = o_p(n^{-1/2}).\]
    for all $a \in \mathcal{A}$.

    Then we show that the first two terms are asymptotic normal. Simple calculation shows that
\[\begin{aligned}
E(X_i|B_i=k) &= \frac{1}{n_{[k]}}\sum\limits_{i=1}^n I(B_i=k)E(X_i|B_i=k) \\
&= \frac{1}{n_{[k]a}} \sum\limits_{i=1}^n I(A_i=a,B_i=k)E(X_i|B_i=k), \\
\bar{X}_{[k]} &= \frac{1}{n_{[k]}}\sum\limits_{i=1}^n I(B_i=k)X_i.  
\end{aligned}\]
Define
\[\begin{aligned}
\mathbb{L}_{n[k],r_i(a)}^{(1)} &= \frac{1}{\sqrt{n}} \sum\limits_{i=1}^n I(A_i=a, B_i=k)\tilde{r}_i(a),  \\
\mathbb{L}_{n[k],X_i}^{(1)} &= \frac{1}{\sqrt{n}} \sum\limits_{i=1}^n I(B_i=k)\tilde{X}_i, \\
\mathbb{L}_{n[k]}^{(2)} &= \sqrt{n}(p_{[k]} - p_{n[k]}).
\end{aligned}\]
Here $r_i(a)$ is a transformed outcome, and
\[\tilde{r}_i(a) = r_i(a) - E[r_i(a)|B_i=k], \quad i \in [k].\]
In \cite{Bugni2019}, it takes $r_i(a) = Y_i(a)$. Some calculation for the first two parts of $\hat{\tau}_{\rm{gen}}$ leads to
\[\begin{aligned}
\sqrt{n} (\hat{\tau}_{\rm{gen}} - \tau) = &\big\{\sum\limits_{k \in \mathcal{K}}(\frac{1}{\pi_{[k]a}}\mathbb{L}^{(1)}_{n[k],r_{i,\rm{gen}}(a)} - \frac{1}{\pi_{[k]0}}\mathbb{L}^{(1)}_{n[k],r_{i,\rm{gen}}(0)}), a\in \mathcal{A}\big\} \\
&+ \big\{\sum\limits_{k \in \mathcal{K}}\mathbb{L}^{(1) \rm T}_{n[k],X_i}\{\beta_{[k]}(a)-\beta_{[k]}(0)\}, a\in \mathcal{A}\big\} \\ 
&+ \big\{\sum\limits_{k \in \mathcal{K}}\mathbb{L}_{n[k]}^{(2)} \{m_{[k]}(a) - m_{[k]}(0)\}, a\in \mathcal{A}\big\} + o_p(1),
\end{aligned}
\]
where $r_{i,\rm{gen}}(a) = Y_i(a) - X_i^{\rm T} \beta_{[k]}(a)$ and $m_{[k]}(a) = E[Y_i(a)\mid B_i=k] - E[Y_i(a)]$.
By Lemma \ref{lem} and some additional calculations, we have
\begin{multline*}
\begin{pmatrix}
\big\{\sum\limits_{k \in \mathcal{K}}(\frac{1}{\pi_{[k]a}}\mathbb{L}^{(1)}_{n[k],r_{i,\rm{gen}}(a)} - \frac{1}{\pi_{[k]0}}\mathbb{L}^{(1)}_{n[k],r_{i,\rm{gen}}(0)}), a\in \mathcal{A}\big\} \\
\big\{\sum\limits_{k \in \mathcal{K}}(\mathbb{L}^{(1) \rm T}_{n[k],X_i}\{\beta_{[k]}(a)-\beta_{[k]}(0)\}, a\in \mathcal{A}\big\}\\
\big\{\sum\limits_{k \in \mathcal{K}}\mathbb{L}_{n[k]}^{(2)}\{ m_{[k]}(a) - m_{[k]}(0)\}, a\in \mathcal{A}\big\}
\end{pmatrix}
\stackrel{d}\longrightarrow \\
\operatorname{N}\left( \begin{pmatrix}
0\\0\\0
\end{pmatrix}, \begin{pmatrix}
V_{\tilde{r}_{\rm{gen}}} & V_{\tilde{r}_{\rm{gen}}\tilde{X}} & 0 \\
V_{\tilde{r}_{\rm{gen}}\tilde{X}}^{\rm T}& V_{\tilde{X}} & 0 \\
0& 0 & V_{H\tilde{Y}}
\end{pmatrix}\right)
,    
\end{multline*}
where 
\begin{align*}
    V_{H\tilde{Y}} &= \sum\limits_{k \in \mathcal{K}} p_{[k]} \left(m_{[k]}(a) - m_{[k]}(0): a \in \mathcal{A}\right) \times  \left(m_{[k]}(a') - m_{[k]}(0): a' \in \mathcal{A}\right)^{\mathrm T}, \\
    V_{\tilde{r}_{\rm{gen}}} & = \sum\limits_{k \in \mathcal{K}}\frac{p_{[k]}}{\pi_{[k]0}} \sigma^2_{[k]\tilde{r}_{\rm{gen}}(0)} 1_A1_A^{\mathrm T} + \diag\left\{\sum\limits_{k \in \mathcal{K}} \frac{p_{[k]}}{\pi_{[k]a}} \sigma^2_{[k]\tilde{r}_{\rm{gen}}(a)}: a\in \mathcal{A}\right\},  \\
    V_{\tilde{X}} &= \Big\{\sum\limits_{k \in \mathcal{K}} p_{[k]}\{\beta_{[k]}(a) - \beta_{[k]}(0)\}^{\mathrm T} \Sigma_{[k]XX} \{\beta_{[k]}(a') - \beta_{[k]}(0)\}: (a,a') \in \mathcal{A}\times \mathcal{A}\Big\},\\
    V_{\tilde{r}_{\rm{gen}}\tilde{X}} & = \Big\{\sum\limits_{k\in\mathcal{K}}p_{[k]}\{\beta_{[k]}(a')-\beta_{[k]}(0)\}^{\mathrm T}\big[(\Sigma_{[k]XY(a)}-\Sigma_{[k]XY(0)})- \Sigma_{[k]XX}\{\beta_{[k]}(a)-\beta_{[k]}(0)\}\big]: \\
    &(a,a')\in \mathcal{A}\times\mathcal{A}\Big\}.
\end{align*} 

Let $V_{\rm{gen}} = V_{H\tilde{Y}} + V_{\tilde{r}_{\rm{gen}}} + V_{\tilde{X}} + V_{\tilde{r}_{\rm{gen}}\tilde{X}} + V_{\tilde{r}_{\rm{gen}}\tilde{X}}^{\rm T}$, we have
\[\sqrt{n} (\hat{\tau}_{\rm{gen}} - \tau) \stackrel{d}\longrightarrow \operatorname{N}(0, V_{\rm{gen}}).\]
\end{proof}

\section{Proof of main theorems}
\subsection{Proof of Proposition 1}
\begin{proof}
    By the definition of Lasso estimator, we have 
    \[\HLassoBeta = \arg\min_{\beta\in\mathbb{R}^p} \frac{1}{n_{[k]a}} \sum\limits_{i=1}^n I(A_i=a, B_i=k) \{Y_i-\bar Y_{[k]a} - (X_i - \bar X_{[k]a})^{\rm T}\beta\}^2 + \lambda_{[k]a}\norm{\beta}{1}\]
    Therefore,  we have
    \begin{multline*}
        \frac{1}{n_{[k]a}} \sum\limits_{i=1}^n I(A_i=a, B_i=k) \{Y_i-\bar Y_{[k]a} - (X_i - \bar X_{[k]a})^{\rm T}\HLassoBeta\}^2 + \lambda_{[k]a}\norm{\HLassoBeta}{1}\\
        \leq \frac{1}{n_{[k]a}} \sum\limits_{i=1}^n I(A_i=a, B_i=k) \{Y_i-\bar Y_{[k]a} - (X_i - \bar X_{[k]a})^{\rm T}\ProjBeta\}^2 + \lambda_{[k]a}\norm{\ProjBeta}{1}. 
    \end{multline*}
    Let $u = \HLassoBeta - \ProjBeta$. Let $\epsilon_i(a) = Y_i(a) - X_i^{\rm T}\ProjBeta$, for $i\in (a,k)$, be a transformed outcome. Here $i\in (a,k)$ means the units such that $A_i=a, B_i=k$. Also, we denote
    \begin{align*}
        S_{[k]XX(a)} = \frac{1}{n_{[k]a}} \sum\limits_{i=1}^n I(A_i=a,B_i=k)(X_i-\bar X_{[k]a})(X_i - \bar X_{[k]a})^{\rm T} \\
        S_{[k]X\epsilon(a)} = \frac{1}{n_{[k]a}} \sum\limits_{i=1}^n I(A_i=a,B_i=k)(X_i-\bar X_{[k]a})(\epsilon_i(a) - \bar \epsilon_{[k]a}).
    \end{align*}
    Then by some calculations,
    \[u^{\rm T}S_{[k]XX(a)}u \leq \lambda_{[k]a}\norm{\ProjBeta}{1} - \lambda_{[k]a}\norm{\HLassoBeta}{1} + 2u^{\rm T}S_{[k]X\epsilon(a)}.\]
    By Lemma \ref{lem2}, it has high probability that 
    \[2\norm{S_{[k]X\epsilon(a)}}{\infty} \leq \frac{\lambda_{[k]a}}{2}.\]
    Therefore, we have 
    \[u^{\rm T}S_{[k]XX(a)}u \leq \lambda_{[k]a}\norm{\ProjBeta}{1} - \lambda_{[k]a}\norm{\HLassoBeta}{1} + \frac{\lambda_{[k]a}}{2}\norm{u}{1}.\]
    As $S_{[k]}$ is the union of the support of $\ProjBeta$ for $a\in\mathcal{A}_0$, we have 
    \[\norm{[\ProjBeta]_{S^C_{[k]}}}{1} = 0.\]
    Therefore, we split the $l_1$ norm into the parts in $S_{[k]}$ and $S^C_{[k]}$, and by triangle inequality: 
    \begin{align*}
        \norm{[[\ProjBeta]_{S_{[k]}}}{1} - \norm{[\HLassoBeta]_{S_{[k]}}}{1} \leq \norm{[u]_{S_{[k]}}}{1}, \\
        \norm{[\HLassoBeta]_{S_{[k]}^C}}{1} \geq \norm{[u]_{S_{[k]}^C}}{1} - \norm{[\ProjBeta]_{S_{[k]}^C}}{1},
    \end{align*}
    we have 
    \[u^{\rm T}S_{[k]XX(a)}u \leq \frac{3}{2}\lambda_{[k]a}\norm{u_{S_{[k]}}}{1} - \frac{1}{2}\lambda_{[k]a}\norm{u_{S^C_{[k]}}}{1}.\]
    As all three terms are greater than or equal to zero, we have $\norm{u_{S^C_{[k]}}}{1} \leq 3\norm{u_{S^C_{[k]}}}{1}$.
    Let 
    \[E_3 = \{u\in V:\ u^{\rm T}S_{[k]XX(a)}u \geq \kappa \norm{u}{2}^2\}\]
    where $V = \{h\in\mathbb{R}^p:\ \norm{h_{S^C_{[k]}}}{1}\leq 3\norm{h_{S_{[k]}}}{1}\}$. By Lemma \ref{lem3}, 
    \[\Pr(E_3)\rightarrow 1.\]
    Under $E_3$, we have 
    \[\kappa\norm{u}{2}^2 \leq \frac{3}{2}\lambda_{[k]a}\norm{u_{S_{[k]}}}{1} \leq\frac{3}{2} \sqrt{s_{[k]}}\lambda_{[k]a}\norm{u_{S_{[k]}}}{2} \leq \frac{3}{2} \sqrt{s_{[k]}}\lambda_{[k]a}\norm{u}{2}.\]
    That is 
    \[\norm{u}{2} \leq \frac{3}{2\kappa} \sqrt{s_{[k]}}\lambda_{[k]a}.\]
    Then 
    \[\norm{u}{1} = \norm{u_{S_{[k]}}}{1} + \norm{u_{S_{[k]}}}{1} \leq 4\norm{u_{S_{[k]}}}{1} \leq 4\sqrt{s_{[k]}}\norm{u}{2} \leq \frac{6}{\kappa}s_{[k]}\lambda_{[k]a}. \]
    Therefore, as $\lambda_{[k]a} = O_p(\sqrt{M_{n[k]}\log p/n})$, we have
    \[\norm{\HLassoBeta - \ProjBeta}{1} = O_p\left(s_{[k]}\sqrt{\frac{M_{n[k]}\log p}{n}}\right).\]
    Then by Lemma \ref{lem1}, $\norm{\bar X_{[k]a} - \bar X_{[k]}}{\infty} = O_p(\sqrt{\log p/n})$, we have 
    \begin{align*}
        &|\sqrt{n}(\bar X_{[k]a} - \bar X_{[k]})^{\rm T}(\HLassoBeta - \ProjBeta)| \\
        &= \sqrt{n}\cdot O_p\left(s_{[k]}\sqrt{\frac{M_{n[k]}\log p}{n}}\right) \cdot  O_p\left(\sqrt{\frac{\log p}{n}}\right)\\
        &= O_p\left(s_{[k]}\log p \sqrt{\frac{M_{n[k]}}{n}}\right)\\
        &=o_p(1).
    \end{align*}
    The last line is due to Assumption \ref{ap4}. The assumptions for Proposition \ref{prop2} hold, so we have 
    \[\sqrt{n}(\hat\tau_{\rm lasso} - \tau) \stackrel{d}\rightarrow \operatorname{N}(0,V_{\rm \rm{gen}}).\]
    In Proposition \ref{prop2}, when $\beta_{[k]}(a) = \ProjBeta$ for $k\in\mathcal{K}, a\in\mathcal{A}_0$, we have
    \[(\Sigma_{[k]XY(a)}-\Sigma_{[k]XY(0)})- \Sigma_{[k]XX}\{\beta_{[k]}(a)-\beta_{[k]}(0)\}=0.\]
    Therefore, the asymptotic covariance matrix is $V_{\rm proj}$ where $V_{\rm proj}=V_{\tilde r_{\rm proj}} + V_{\tilde X} + V_{H\tilde Y}.$
\end{proof}

\subsection{Proof of Theorem 1}
\begin{proof}\ 

\noindent \textbf{Transfer Step.}

    By the definition of $\ProjBetaSourceE$, we have 
    \begin{multline*}
        \frac{1}{\source n_{[k]a}}\sum\limits_{i\in (a,k)} \big\{\source{Y_i} - \source{\bar Y_{[k]a}} - (\source{X_i} - \source{\bar X_{[k]a}})^{\rm T}\ProjBetaSourceE\big\}^2 + \source \lambda_{[k]a}\norm{\ProjBetaSourceE}{1}\\
       \leq \frac{1}{\source n_{[k]a}}\sum\limits_{i\in (a,k)} \big\{\source{Y_i} - \source{\bar Y_{[k]a}} - (\source{X_i} - \source{\bar X_{[k]a}})^{\rm T}\ProjBetaSource\big\}^2 + \source \lambda_{[k]a} \norm{\ProjBetaSource}{1}.
    \end{multline*} 
     Let $\source{u} = \ProjBetaSourceE - \ProjBetaSource$. By some calculations, we have 
    \[(\source{u})^{\rm T}\source{S_{[k]XX(a)}}\source{u} \leq \source\lambda_{[k]a} \lVert \ProjBetaSource\rVert_1-\source\lambda_{[k]a} \lVert \ProjBetaSourceE\rVert_1+ 2(\source{u})^{\rm T} \source S_{[k]X\epsilon(a)}\]
    where
    \[\source{S_{[k]XX(a)}} = \frac{1}{n_{[k]a}}\sum\limits_{i=1}^n I(\source{A_i}=a,\source{B_i}=k)(\source{X_i} - \source{\bar X_{[k]a}})(\source{X_i} - \source{\bar X_{[k]a}})^{\rm T},\]
    \[\source S_{[k]X\epsilon(a)} = \frac{1}{n_{[k]a}}\sum\limits_{i=1}^n I(\source{A_i}=a,\source{B_i}=k)(\source{X_i} - \source{\bar X_{[k]a}})(\source \epsilon_i(a) - \source{\bar \epsilon_{[k]a}}).\]
    Here $\source{\epsilon_i}(a) = \source Y_i(a) - \source X_i \source\ProjBeta$, for $i \in (a,k)$. 
    
    Let $E_1 = \{\norm{2\source S_{[k]X\epsilon(a)}}{\infty}\leq \source\lambda_{[k]a}/2\}$. By Lemma \ref{lem2}, we have
    \[\Pr(E_1)\geq 1-\frac{1}{M_{n[k]}} \rightarrow 1.\]
    Then under $E_1$,
    \[(\source{u})^{\rm T}\source{S_{[k]XX(a)}}\source{u} \leq \source\lambda_{[k]a} \lVert \ProjBetaSource\rVert_1-\source\lambda_{[k]a} \lVert \ProjBetaSourceE\rVert_1+ \frac{\source\lambda_{[k]a}}{2}\norm{\source{u}}{1}.\]
    Recall that $S_{[k]} = \bigcup_{a\in\mathcal{A}}\{j\in\{1,\dots,p\}:\ \beta_{j,[k]\rm proj}(a)\neq 0\}$ and $S^C_{[k]}$ is the complementary set of $S_{[k]}$. Let $[u]_S = (u_j, j\in S_{[k]})$ and $[u]_{S^C} = (u_j, j\in S^C_{[k]})$ for any vector $u$.  We split $S_{[k]}$ and $S_{[k]}^C$ apart and take following inequalities: 
\[\norm{[\ProjBetaSource]_S}{1} - \norm{[\ProjBetaSourceE]_{S}}{1} \leq \norm{[\source{u}]_S}{1},\]
\[ \norm{[\source{u}]_{S^C}}{1}\leq \norm{[\ProjBetaSourceE]_{S^C}}{1} + \norm{[\ProjBetaSource]_{S^C}}{1}.\]
Then we have 
    \[(\source{u})^{\rm T}\source{S_{[k]XX(a)}}\source{u} \leq \frac{3}{2}\source\lambda_{[k]a}\norm{[\source{u}]_S}{1} + \frac{3}{2}\source\lambda_{[k]a} \lVert [\ProjBetaSource]_{S^C}\rVert_1-\frac{1}{2}\source\lambda_{[k]a} \lVert [\source{u}]_{S^C}\rVert_1.\]
Then we compare the first term and the second term on the RHS.

\noindent \textbf{Case 1 : \[\frac{3}{2}\source\lambda_{[k]a}\norm{[\source{u}]_S}{1} \geq \frac{3}{2}\source\lambda_{[k]a} \lVert [\ProjBetaSource]_{S^C}\rVert_1.\]}
Then we have $\norm{[\source{u}]_{S^C}}{1}\leq 6 \norm{[\source{u}]_S}{1}$. By Lemma \ref{lem3}, we have 
\[c_0 \norm{\source{u}}{2}^2 \leq 3 \source\lambda_{[k]a}\norm{[\source{u}]_S}{1} \leq 3\sqrt{s_{[k]}} \source\lambda_{[k]a}\norm{[\source{u}]_S}{2},\]
which leads to 
\[\norm{\source{u}}{2} \leq C_1\sqrt{s_{[k]}} \source\lambda_{[k]a} \]
and
\[\norm{\source{u}}{1} \leq 7\norm{[\source{u}]_S}{1} \leq 7\sqrt{s_{[k]}}\norm{[\source{u}]_S}{2}\leq C_2 s_{[k]} \source\lambda_{[k]a}.\]

\noindent \textbf{Case 2 : \[\frac{3}{2}\source\lambda_{[k]a}\norm{[\source{u}]_S}{1} < \frac{3}{2}\source\lambda_{[k]a} \lVert [\ProjBetaSource]_{S^C}\rVert_1.\]}
Then
\[(\source{u})^{\rm T}\source{S_{[k]XX(a)}}\source{u} \leq 3\source\lambda_{[k]a} \lVert [\ProjBetaSource]_{S^C}\rVert_1-\frac{1}{2}\source\lambda_{[k]a} \lVert [\source{u}]_{S^C}\rVert_1.\]
A direct bound for $\norm{\source{u}}{1}$ is 
\[\norm{\source{u}}{1} \leq \norm{[\source{u}]_S}{1} + \lVert [\source{u}]_{S^C}\rVert_1 \leq 7 \norm{[\ProjBetaSource]_{S^C}}{1} \leq 7\norm{[\delta_{[k]}(a)]_{S^C}}{1}\leq 7h.\]
The third inequality holds as 
\[\norm{[\ProjBetaSource]_{S^C}}{1} \leq \norm{[\ProjBeta]_{S^C}}{1} + \norm{[\delta_{[k]}(a)]_{S^C}}{1} = \norm{[\delta_{[k]}(a)]_{S^C}}{1}.\]
Also, by \textcolor{red}{RSC} condition and Lemma \ref{lem4}, we have
\[\kappa_1 \norm{\source{u}}{2}^2 - \kappa_3 \frac{\log p}{\source n_{[k]a}}\norm{\source{u}}{1}^2 \leq 3\source\lambda_{[k]a} \lVert [\ProjBetaSource]_{S^C}\rVert_1-\frac{1}{2}\source\lambda_{[k]a} \lVert [\source{u}]_{S^C}\rVert_1.\]
Under condition that $h(\log p/\source n)^{1/2}=o(1)$, we have
\[\kappa_1 \norm{\source{u}}{2}^2 \leq 3\source\lambda_{[k]a} \lVert [\ProjBetaSource]_{S^C}\rVert_1 \leq C_3 \source{\lambda_{[k]a}} h.\]
This is used for the proof in the debiasing step.
Therefore, we have
\[\norm{\source{u}}{1} \leq C_2 s_{[k]} \source\lambda_{[k]a}+C_4 h\]
\noindent \textit{\textbf{Debiasing Step:}}\ 

Let $u = \hat\delta_{[k]}(a) - \delta_{[k]}(a)$.
By similar calculations in the Transfer step, we have

\begin{multline*}
    u^{\rm T}S_{[k]XX(a)}u \leq \lambda_{[k]a} \lVert \delta_{[k]}(a)\rVert_1-\lambda_{[k]a} \lVert \hat\delta_{[k]}(a)\rVert_1 \\
    + 2u^{\rm T}\left[\frac{1}{n_{[k]a}}\sum\limits_{i\in (a,k)} (X_i - \bar X_{[k]a})\{Y_i-\bar Y_{[k]a} - (X_i - \bar X_{[k]a})(\ProjBetaSourceE+\delta_{[k]}(a))\}\right].
\end{multline*}
Recall that  $\epsilon_i(a) = Y_i(a) - X_i^{\rm T} \ProjBeta$, as \[\ProjBeta = \source \ProjBeta + \delta_{[k]}(a) = \delta_{[k]}(a) + \ProjBetaSourceE-\source u,\]
we have
\[u^{\rm T}S_{[k]XX(a)}u \leq \lambda_{[k]a} \lVert \delta_{[k]}(a)\rVert_1-\lambda_{[k]a} \lVert \hat\delta_{[k]}(a)\rVert_1 + 2(u^{\rm T}S_{[k]X\epsilon(a)} - u^{\rm T}S_{[k]XX(a)}\source u).\]
Here $S_{[k]XX(a)}$ and $S_{[k]X\epsilon(a)}$ are the same as the ones defined in the Proof of Proposition 1.
We focus on the third term on the RHS first. By triangle inequality and basic inequality $|ab|\leq \frac{ca^2}{2} + \frac{b^2}{2c}$, we have
\[2|u^{\rm T} S_{[k]X\epsilon(a)} - u^{\rm T} S_{[k]XX(a)} \source{u}|\leq 2|u^{\rm T} S_{[k]X\epsilon(a)}| + \frac{1}{2}u^{\rm T}S_{[k]XX(a)}u + 2(\source{u})^{\rm T}S_{[k]XX(a)}(\source{u}).\]
Let \[E_2 = \left\{\norm{2S_{[k]X\epsilon(a)}}{\infty}\leq \frac{\lambda_{[k]a}}{2}\right\} \cap E_1.\]
By Lemma \ref{lem2} we have 
\[\Pr(E_2) \geq 1-\frac{1}{M_{n[k]}}\rightarrow 1.\]
Therefore, under $E_2$, we have
\[\frac{1}{2}u^{\rm T}S_{[k]XX(a)}u \leq \lambda_{[k]a} \lVert \delta_{[k]}(a)\rVert_1-\lambda_{[k]a} \lVert \hat\delta_{[k]}(a)\rVert_1+ \frac{1}{2}\lambda_{[k]a}\norm{u}{1} + 2(\source{u})^{\rm T}S_{[k]XX(a)}(\source{u}).\]
As 
\[\norm{\hat\delta_{[k]}(a)}{1} \geq \norm{u}{1} - \norm{\delta_{[k]}(a)}{1},\]
we have
\[\frac{1}{2}u^{\rm T}S_{[k]XX(a)}u \leq 2\lambda_{[k]a} \lVert \delta_{[k]}(a)\rVert_1- \frac{1}{2}\lambda_{[k]a}\norm{u}{1} + 2(\source{u})^{\rm T}S_{[k]XX(a)}(\source{u}).\]

\noindent \textbf{Case 1: \[\lambda_{[k]a} \lVert \delta_{[k]}(a)\rVert_1 \geq (\source{u})^{\rm T}S_{[k]XX(a)}(\source{u}).\]}
We have
\[\norm{u}{1} \leq 8 \norm{\delta_{[k]}(a)}{1} \leq 6h.\]

\noindent \textbf{Case 2: \[\lambda_{[k]a} \lVert \delta_{[k]}(a)\rVert_1 \leq (\source{u})^{\rm T}S_{[k]XX(a)}(\source{u}).\]}
We have 
\[\lambda_{[k]a}\norm{u}{1} \leq 8(\source{u})^{\rm T}S_{[k]XX(a)}(\source{u})\leq C_5 \norm{\source{u}}{2}^2.\]
The second inequality holds as $X_i$ is uniformly bounded and 
\[\frac{1}{n_{[k]a}}\sum\limits_{i=1}^n I(A_i=a,B_i=k) \big[(X_i-\bar X_{[k]a})^{\rm T}\source u\big]^2 \leq 4M^2\norm{\source u}{2}^2. \]
In the Transfer Step, we already have 
\[\norm{\source{u}}{2}^2 \leq C_1^2 s_{[k]}(\source{\lambda}_{[k]a})^2 + C_3\source{\lambda}_{[k]a}h.\]
Therefore,
\[\norm{u}{1} \leq \frac{C_5}{\lambda_{[k]a}}\left(C_1^2 s_{[k]}(\source{\lambda}_{[k]a})^2 + C_3\source{\lambda}_{[k]a}h\right).\]
By Assumptions \ref{ap4} and \ref{ap7}, we have
\[\source{\lambda}_{[k]a} = O\left(\sqrt{\frac{M_{\source n_{[k]}}\log p}{\source n}}\right),\]
\[\lambda_{[k]a} = O\left(\sqrt{\frac{M_{n[k]}\log p}{n}}\right).\]
We assume that $M_{N}$ is the same and consider the convergence of the subsequence. We can require that $M_N$ is monotonically increasing but $M_N/N$ is decreasing when $N$ is large enough. $M_N$ can diverge slowly to satisfy our conditions, such as $M_N = \log\log N$. Then 
\[\frac{\source{\lambda}_{[k]a}}{\lambda_{[k]a}} \leq 1\]when $\source n$ and $n$ are large enough.
Therefore,
\[\norm{u}{1}\leq K_1 s_{[k]}\source{\lambda}_{[k]a} + K_2h.\]
According to the inequalities for $\norm{\source u}{1}$ and $\norm{u}{1}$, in high probability that we have
\begin{align*}
    \norm{\hat\beta_{[k]\rm tl}(a) - \ProjBeta}{1}  &=  \norm{\{\ProjBetaSourceE -\ProjBetaSource\} + \{ \hat\delta_{[k]}(a)- \delta_{[k]}(a)\}}{1} \\  
    &\leq \norm{\source u}{1} + \norm{u}{1} \\
    &\leq K_1 s_{[k]}\sqrt{\frac{\source{M}_{n[k]}\log p}{\source n}} + K_2h
\end{align*}
for some constant $K_1$ and $K_2$ independent with $n,\source n$ and $ p$. When $h<s_{[k]}\sqrt{\log p/n}$, we can write the above term as
\[ \norm{\hat\beta_{[k]\rm tl}(a) - \ProjBeta}{1}\leq K_1 s_{[k]}\sqrt{\frac{\source M_{n[k]}\log p}{\source n+n}} + K_2h\wedge s_{[k]}\sqrt{\frac{M_{n[k]}\log p}{n}}.\]
\end{proof}

\subsection{Proof of Theorem 2}

According to Theorem \ref{cor1} and Lemma \ref{lem1}, we have 
    \[\norm{\hat\beta_{[k]\rm tl}(a) - \ProjBeta}{1} = O_p\left(s_{[k]}\sqrt{\frac{\source M_{n[k]}\log p}{n+\source n}} + h\right),\]
    and 
\[\norm{\bar X_{[k]a} - \bar X_{[k]}}{\infty} = O_p\left\{\left(\frac{\log p}{n}\right)^{1/2}\right\}\]
for all $k\in\mathcal{K}$ and $a\in\mathcal{A}_0$. By H$\ddot{\rm o}$lder's inequality, we have 
\begin{align*}
    |\sqrt{n}(\bar X_{[k]a} - \bar X_{[k]})^{\rm T}&\{\hat\beta_{[k]\rm tl}(a) - \ProjBeta\}| \\
    &\leq \sqrt{n}\norm{\bar X_{[k]a} - \bar X_{[k]}}{\infty}\norm{\hat\beta_{[k]\rm tl}(a) - \ProjBeta}{1} \\
    &= \sqrt{n}\cdot O_p\left(s_{[k]}\sqrt{\frac{\source M_{n[k]}\log p}{n+\source n}} + h\right)\cdot O_p\left\{\left(\frac{\log p}{n}\right)^{1/2}\right\} \\
    &= O_p\left(s_{[k]}\log p \sqrt{\frac{M_n}{n+\source n}} + h\sqrt{\log p}\right) \\
    &=o_p(1)
\end{align*} 
The last line is due to Assumption \ref{ap7} and conditions that $h(\log p)^{1/2} = o(1)$. Therefore, by Proposition \ref{prop2} and similar arguments for the asymptotic covariance matrix in the proof of Proposition \ref{prop1}, 
\[\sqrt{n}(\hat\tau_{\rm tl} - \tau) \stackrel{d}\longrightarrow \operatorname{N}(0,V_{\rm proj}).\]
\subsection{Proof of Theorem 3}
    By Lemma \ref{lem5}, we only need to show that 
    \[\frac{1}{n_{[k]}}\sum\limits_{i\in [k]} [X_i^{\rm T}\{\hat\beta_{[k]\rm lasso}(a) - \ProjBeta\}]^2 = o_p(1)\]
    and
    \[\frac{1}{n_{[k]}}\sum\limits_{i\in [k]} [X_i^{\rm T}\{\hat\beta_{[k]\rm tl}(a) - \ProjBeta\}]^2 = o_p(1),\]
    respectively.
    By Proposition \ref{prop1} and Theorem \ref{thm1}, 
    \[\norm{\hat\beta_{[k]\rm lasso}(a) - \ProjBeta}{1} = O_p\left(s_{[k]}\sqrt{\frac{M_n\log p}{n}}\right) = o_p(1)\]
    and 
    \[\norm{\hat\beta_{[k]\rm lasso}(a) - \ProjBeta}{1} = O_p\left(s_{[k]}\sqrt{\frac{\source M_n\log p}{n+\source n}}+h \right) = o_p(1).\]
    By H$\ddot{\rm o}$lder inequality and $X_i$ is uniformly bounded, we have 
    \[\frac{1}{n_{[k]}}\sum\limits_{i\in [k]} [X_i^{\rm T}\{\hat\beta_{[k]\rm lasso}(a) - \ProjBeta\}]^2 \leq M^2 \norm{\hat\beta_{[k]\rm lasso}(a) - \ProjBeta}{1}^2 = o_p(1)\]
    and 
    \[\frac{1}{n_{[k]}}\sum\limits_{i\in [k]} [X_i^{\rm T}\{\hat\beta_{[k]\rm lasso}(a) - \ProjBeta\}]^2 \leq M^2 \norm{\hat\beta_{[k]\rm tl}(a) - \ProjBeta}{1}^2 = o_p(1).\]
    Therefore our variance estimators are consistent.
\subsection{Lemmas with proof}

\begin{lemma}\label{lem}
Let $\mathbb{L}_n^{(2)} =\{\mathbb{L}_{n[k]}^{(2)}, k\in \mathcal{K} \}$, and 
\[\mathbb{L}_n = \Bigg(\bigg\{\sum\limits_{k \in \mathcal{K}} \frac{1}{\pi_{[k]a}} \mathbb{L}^{(1)}_{n[k],r_{i,\rm{gen}}(a)}, a\in\mathcal{A}_0\bigg\},\bigg\{\sum\limits_{k \in \mathcal{K}} \mathbb{L}^{(1) \rm T}_{n[k],X_i}\beta_{[k]}(a), a\in \mathcal{A}_0\bigg\}, \mathbb{L}^{(2)}_{n}\Bigg).\] 
Under Assumptions \ref{ap1}--\ref{ap2}, we have
\[\mathbb{L}_n \stackrel{d}\longrightarrow \operatorname{N}\left( \begin{pmatrix}
0\\0\\0
\end{pmatrix}, 
\begin{pmatrix}
\Sigma_1 &\Sigma_{12} &0 \\
\Sigma_{12}^{\rm T} & \Sigma_2 & 0 \\
0 & 0 & \Sigma_3 \\
\end{pmatrix}\right),\]
where
\[\begin{aligned}
\Sigma_1 &= \diag\Bigg\{\sum\limits_{k\in\mathcal{K}} \frac{p_{[k]}}{\pi_{[k]a}} \sigma^2_{[k]\tilde{r}_{\rm{gen}}(a)}, a \in \mathcal{A}_0\Bigg\}, \\
\Sigma_{12} &= \Bigg\{\sum\limits_{k \in \mathcal{K}} p_{[k]} \beta_{[k]}^{\rm T}(a')\Bigg(\Sigma_{[k]XY(a)} - \Sigma_{[k]XX}\beta_{[k]}(a)\Bigg), (a,a') \in \mathcal{A}_0 \times \mathcal{A}_0\Bigg\},\\
\Sigma_2 &= \Bigg\{\sum\limits_{k\in\mathcal{K}} \beta_{[k]}^{\rm T}(a) \Sigma_{\tilde{X}\tilde{X}} \beta_{[k]}(a'), (a,a') \in \mathcal{A}_0\times \mathcal{A}_0\Bigg\}, \\
\Sigma_3 &= \diag\{p_{[k]}: k \in \mathcal{K}\} -\{p_{[k]}p_{[k']}, (k,k')\in \mathcal{K} \times \mathcal{K}\}.
\end{aligned}\]

\end{lemma}
\begin{proof}
    Most of the proof is similar to the proof for Lemma 5 in \cite{gu2023regression}. The difference is that in this Lemma, $\beta_{[k]}(a)$ may not be $\ProjBeta$ as we define in the context. The consequence is that $\Sigma_{12} \neq 0$ for most cases. The remaining parts are the same and we omit the proof.
\end{proof}
\begin{lemma}\label{lem1}
    Under Assumptions \ref{ap1} and \ref{ap2}, we have 
    \[\norm{\bar X_{[k]a} - \bar X_{[k]}}{\infty} = O_p\left\{\left(\frac{\log p}{n}\right)^{1/2}\right\}.\]
    for all $a \in \mathcal{A}_0,\ k \in \mathcal{K}$.
\end{lemma}
\begin{proof}
    The proof is similar to the proof of Lemma S4 in \cite{liu2022lasso}. We can replace $A_i=1$ into $A_i=a$ for all $a \in\mathcal{A}_0$ to show our result. Therefore, we omit the proof.
\end{proof}
\begin{lemma}\label{lem2}
    Under Assumptions \ref{ap1}-\ref{ap4} and $E\{\epsilon^2_i(a)\} < \infty$ for all $a \in \mathcal{A}_0$, the events 
    \[E_1 = \left\{\norm{2\source{S_{[k]X\epsilon(a)}}}{\infty} \leq \frac{\source \lambda_{[k]a}}{2}\right\},\quad E'_2 = \left\{\norm{2S_{[k]X\epsilon(a)}}{\infty} \leq \frac{\lambda_{[k]a}}{2}\right\}\]
    have probability one when source size $\source n$ and target size $n$ goes to infinity.
\end{lemma}

\begin{proof}
     The proof is similar to the proof of Lemma S5 in \cite{liu2022lasso}. It's a special case for stratum size $K=1$. At the same time, we can replace $A_i=1$ into $A_i=a$ for all $a \in\mathcal{A}_0$ to show our result. Therefore, we omit the proof.
\end{proof}

\begin{lemma}\label{lem3}
        Under Assumptions \ref{ap1}--\ref{ap4}, for any $u \in V = \{h\in\mathbb{R}^p:\ \norm{h_{S^C_{[k]}}}{1}\leq 3\norm{h_{S_{[k]}}}{1}\}$, we have
        \[u^{\rm T}S_{[k]XX(a)}u \geq \kappa \norm{u}{2}\]
        with probability $1-c_1\exp(-c_2n)$, for some positive constant $\kappa$ independent of $n$. 
\end{lemma}

\begin{proof}
    We use a similar argument in \citet{Bugni2018}. As $\{Y_i(0),\dots,Y_i(A),B_i,X_i\}$ are i.i.d. data and $A^{(n)}$ is independent with $\{Y_i(0),\dots,Y_i(A),X_i\}$ conditional on $B^{(n)}$. Therefore, given the $A^{(n)}$ and $B^{(n)}$, the distribution of $S_{[k]XX(a)}$ is the same if we order the units first by strata and then by treatments within each strata. Then, independently with $k = 1,2,\dots,K$ and $(A^{(n)},B^{(n)})$, let $\{Y^k_i(0),\dots,Y_i^k(A),X_i^k\}$ has the same distribution with $\{Y_i(0),\dots,Y_i(A),X_i\}$ given $B_i=k$. Then $S_{[k]XX(a)}| A^{(n)},B^{(n)}$ has the same distribution with 
    \[\frac{1}{n_{[k]a}}\sum\limits_{i=1}^{n_{[k]a}} (X^k_i - \frac{1}{n_{[k]a}}\sum\limits_{i=1}^{n_{[k]a}} X^k_i)(X^k_i - \frac{1}{n_{[k]a}}\sum\limits_{i=1}^{n_{[k]a}} X^k_i)^{\rm T}.\]
    By Assumption \ref{ap1}, $X_i$ is uniformly bounded by a constant $M$. As $X_i^k$ has the same distribution with $X_i$ given $B_i=k$, $X_i^k$ is uniformly bounded by constant $M$ so that is a sub-Gaussian random vector. According to Proposition 2 in \citet{negahban2012unified} for generalized linear models, it can be applied directly to the standard linear model. For any $N\geq 1$, we have 
    \begin{align*}
        &h^{\rm T}\left\{\frac{1}{N}\sum\limits_{i=1}^{N} (X^k_i - \frac{1}{N}\sum\limits_{i=1}^{N} X^k_i)(X^k_i - \frac{1}{N}\sum\limits_{i=1}^{N} X^k_i)^{\rm T}\right\}h 
        \\ &\geq\kappa_1 \norm{h}{2}\left\{\norm{h}{2} - \kappa_2 \sqrt{\frac{\log p}{N}}\norm{h}{1} \right\},\quad \mathrm{for\ all}\  \norm{h}{2}\leq 1
    \end{align*}
    with probability at least $1-c_1\exp(-c_2N)$. Here $\kappa_1,\kappa_2$ only depends on the minimum eigenvalue of covariance matrix $\Sigma_{[k]XX}$ and $M$.

As we can divide $\norm{h}{2}$ on both sides, we have 
    \begin{align*}
        &h^{\rm T}\left\{\frac{1}{N}\sum\limits_{i=1}^{N} (X^k_i - \frac{1}{N}\sum\limits_{i=1}^{N} X^k_i)(X^k_i - \frac{1}{N}\sum\limits_{i=1}^{N} X^k_i)^{\rm T}\right\}h 
        \\ &\geq\kappa_1 \norm{h}{2}\left\{\norm{h}{2} - \kappa_2 \sqrt{\frac{\log p}{N}}\norm{h}{1} \right\}.
    \end{align*}
    Next, we consider the vectors in the convex cone
    \[V = \{h\in\mathbb{R}^p:\ \norm{h_{S^C_{[k]}}}{1}\leq 3\norm{h_{S_{[k]}}}{1}\}.\]
    For any vector $h\in V$,
    \[\norm{h}{1} = \norm{h_{S_{[k]}}}{1} + \norm{h_{S_{[k]}^C}}{1} \leq 4\norm{h_{S_{[k]}}}{1} \leq 4\sqrt{s_{[k]}} \norm{h_{S_{[k]}}}{2}.\]
    Therefore, 
    \begin{align*}
        h^{\rm T}&\left\{\frac{1}{N}\sum\limits_{i=1}^{N} (X^k_i - \frac{1}{N}\sum\limits_{i=1}^{N} X^k_i)(X^k_i - \frac{1}{N}\sum\limits_{i=1}^{N} X^k_i)^{\rm T}\right\}h 
        \\ &\geq\kappa_1 \norm{h}{2}\left\{\norm{h}{2} - \kappa_2 \sqrt{\frac{\log p}{N}}\norm{h}{1} \right\}
        \\ &\geq \left(1-4\kappa_2\sqrt{\frac{s_{[k]}\log p}{N}} \right)\kappa_1\norm{h}{2}^2.
    \end{align*}
    Under Assumption $M_{n[k]} s^2_{[k]} \log^2 p/n \rightarrow 0$, we have 
    \[1-4\kappa_2 \sqrt{\frac{s_{[k]}\log p}{n_{[k]}}} <\frac{1}{2}\]
    when $n\rightarrow\infty$.
    Therefore,
    \begin{equation}\label{eq1}
        \Pr\left(h^{\rm T}\left\{\frac{1}{N}\sum\limits_{i=1}^{N} (X^k_i - \frac{1}{N}\sum\limits_{i=1}^{N} X^k_i)(X^k_i - \frac{1}{N}\sum\limits_{i=1}^{N} X^k_i)^{\rm T}\right\}h\geq \frac{1}{2}\kappa_1\norm{h}{2}^2\right)  \rightarrow 1,
    \end{equation}
    when $N\rightarrow\infty$.
    Using the almost sure representation theorem, we can construct $\tilde n_{[k]a}$ (independent of $\{Y^k_i(0),\dots,Y_i^k(A),X_i^k\}$) such that $\tilde n_{[k]a}/n$ has the same distribution with $n_{[k]a}/n$ and $\tilde n_{[k]a}/n \rightarrow \pi_{[k]a}p_{[k]}$ almost surely. Therefore, for any $h\in V$,
    \begin{align*}
        &\Pr\left(h^{\rm T}\left\{\frac{1}{n_{[k]a}}\sum\limits_{i=1}^{n_{[k]a}} (X^k_i - \frac{1}{n_{[k]a}}\sum\limits_{i=1}^{n_{[k]a}} X^k_i)(X^k_i - \frac{1}{n_{[k]a}}\sum\limits_{i=1}^{n_{[k]a}} X^k_i)^{\rm T}\right\}h\geq \frac{1}{2}\kappa_1\norm{h}{2}^2\ \Big | \ A^{(n)},B^{(n)}\right) \\
        &= \Pr\left(h^{\rm T}\left\{\frac{1}{n\frac{\tilde n_{[k]a}}{n}}\sum\limits_{i=1}^{n\frac{\tilde n_{[k]a}}{n}} (X^k_i - \frac{1}{n\frac{\tilde n_{[k]a}}{n}}\sum\limits_{i=1}^{n\frac{\tilde n_{[k]a}}{n}} X^k_i)(X^k_i - \frac{1}{n\frac{\tilde n_{[k]a}}{n}}\sum\limits_{i=1}^{n\frac{\tilde n_{[k]a}}{n}} X^k_i)^{\rm T}\right\}h\geq \frac{1}{2}\kappa_1\norm{h}{2}^2\ \Big | \ A^{(n)},B^{(n)}\right) \\
        &=E\Bigg[\Pr\Bigg(h^{\rm T}\left\{\frac{1}{n\frac{\tilde n_{[k]a}}{n}}\sum\limits_{i=1}^{n\frac{\tilde n_{[k]a}}{n}} (X^k_i - \frac{1}{n\frac{\tilde n_{[k]a}}{n}}\sum\limits_{i=1}^{n\frac{\tilde n_{[k]a}}{n}} X^k_i)(X^k_i - \frac{1}{n\frac{\tilde n_{[k]a}}{n}}\sum\limits_{i=1}^{n\frac{\tilde n_{[k]a}}{n}} X^k_i)^{\rm T}\right\}h\geq \frac{1}{2}\kappa_1\norm{h}{2}^2 \\
        &\Big | \ A^{(n)},B^{(n)},\frac{\tilde n_{[k]a}}{n}\Bigg)\Bigg] \rightarrow 1.    
    \end{align*}
    where the convergence follows from the dominant convergence theorem, $n(\tilde{n}_{[k]a}/n)\rightarrow \infty$ a.s., independence of $\tilde n_{[k]a}/n$ and $\{Y_i(0)^k,\dots,Y^k_i(A),X^k_i\}$ and (\ref{eq1}).
    Therefore 
    \begin{align*}
        &\Pr \left(h^{\rm T} S_{[k]XX(a)} h \geq \frac{1}{2}\kappa_1 \norm{h}{2}^2\right) \\
        & = \Pr\left(h^{\rm T}\left\{\frac{1}{n_{[k]a}}\sum\limits_{i=1}^{n_{[k]a}} (X^k_i - \frac{1}{n_{[k]a}}\sum\limits_{i=1}^{n_{[k]a}} X^k_i)(X^k_i - \frac{1}{n_{[k]a}}\sum\limits_{i=1}^{n_{[k]a}} X^k_i)^{\rm T}\right\}h\geq \frac{1}{2}\kappa_1\norm{h}{2}^2\right) \\
        &\rightarrow 1.
    \end{align*}
\end{proof}

    \begin{lemma}\label{lem4}
         Under Assumptions \ref{ap1}--\ref{ap2} and \ref{ap5}, we have
        \[u^{\rm T}S_{[k]XX(a)}u \geq \kappa_1 \norm{u}{2}^2 - \kappa_3\frac{\log p}{n}\norm{u}{1}^2\]
        with probability $1-c_1\exp(-c_2n)$, where $\kappa_1,\kappa_3,c_1,c_2$ are independent of $n$.
    \end{lemma}
    \begin{proof}
        Similar to the proof of Lemma \ref{lem3}, we construct $\{Y_i^k(0),\dots,Y_i^k(A),X_i^k\}$ and $S_{[k]XX(a)}$ has the same distribution with 
    \[\frac{1}{n_{[k]a}}\sum\limits_{i=1}^{n_{[k]a}} (X^k_i - \frac{1}{n_{[k]a}}\sum\limits_{i=1}^{n_{[k]a}} X^k_i)(X^k_i - \frac{1}{n_{[k]a}}\sum\limits_{i=1}^{n_{[k]a}} X^k_i)^{\rm T}.\]
    Also, by the same argument in \citet{negahban2012unified}, when $N\rightarrow \infty$,
        \begin{align*}
        &h^{\rm T}\left\{\frac{1}{N}\sum\limits_{i=1}^{N} (X^k_i - \frac{1}{N}\sum\limits_{i=1}^{N} X^k_i)(X^k_i - \frac{1}{N}\sum\limits_{i=1}^{N} X^k_i)^{\rm T}\right\}h 
        \\ &\geq\kappa_1 \norm{h}{2}\left\{\norm{h}{2} - \kappa_2 \sqrt{\frac{\log p}{N}}\norm{h}{1} \right\}.
    \end{align*}
    By a general inequality, we have
    \[\kappa_1 \kappa_2 \sqrt{\frac{\log p}{N}} \norm{h}{1}\norm{h}{2} \leq \frac{\alpha}{2} \norm{h}{2}^2 + \frac{(\kappa_1 \kappa_2)^2}{2\alpha} \frac{\log p}{N} \norm{h}{1}^2.\]
    for some positive constant $\alpha$. We can choose $\alpha = \kappa_1$, then we have
    \begin{align*}
        &h^{\rm T}\left\{\frac{1}{N}\sum\limits_{i=1}^{N} (X^k_i - \frac{1}{N}\sum\limits_{i=1}^{N} X^k_i)(X^k_i - \frac{1}{N}\sum\limits_{i=1}^{N} X^k_i)^{\rm T}\right\}h 
        \\ &\geq\frac{\kappa_1}{2} \norm{h}{2}^2 -\kappa_3 \frac{\log p}{N}\norm{h}{1}^2.
    \end{align*}
    for some positive constant $\kappa_1$ and $\kappa_3$. Then by the similar arguments in the proof of Lemma \ref{lem3}, we have 
    \[\Pr\left(h^{\rm T} S_{[k]XX(a)} h \geq \frac{\kappa_1}{2} \norm{h}{2}^2 -\kappa_3 \frac{\log p}{N}\norm{h}{1}^2 \right) \rightarrow 1.\]
    \end{proof}

\begin{lemma}\label{lem5}
    Under conditions in Proposition \ref{prop2}, suppose there exist $\hat\beta_{[k]}(a)$ and $\beta_{[k]}(a)$, such that
    \[\frac{1}{n_{[k]}}\sum\limits_{i\in [k]} [X_i^{\rm T}\{\hat\beta_{[k]}(a) - \beta_{[k]}(a)\}]^2 = o_p(1)\]
    for all $k\in\mathcal{K}, a\in\mathcal{A}_0$. Let $\hat\sigma^2_{\rm{gen},bc}$ be the estimator of $\sigma^2_{\rm{gen},bc}$, such that 
    \[\sigma_{\rm{gen},bc}^2 = e^{\mathrm T}_{bc}V_{\rm{gen}}e_{bc} = V_{\rm{gen}}(b,b) - 2V_{\rm{gen}}(b,c)+V_{\rm{gen}}(c,c).\]
    Then we have 
    \begin{align*}
\hat{\sigma}^2_{\rm{gen},bc} &= \hat{\sigma}^2_{r_{\rm{gen}}(b)} + \hat{\sigma}^2_{r_{\rm{gen}}(c)} + \hat{\sigma}^2_{bc,HY} \\
&- \sum\limits_{k \in \mathcal{K}}p_{n[k]}\big\{\hat{\beta}_{[k]}(b)-\hat{\beta}_{[k]}(c)\big\}^{\mathrm T} S_{[k]XX}\big\{\hat{\beta}_{[k]}(b)-\hat{\beta}_{[k]}(c)\big\} \\
    &+ 2\sum\limits_{k\in\mathcal{K}}p_{n[k]}\big\{\hat{\beta}_{[k]}(b)-\hat{\beta}_{[k]}(c)\big\}^{\mathrm T}\left\{(S_{[k]XY(b)}-S_{[k]XY(c)})\right\}
    \end{align*}
    is a consistent estimator for $\sigma^2_{\rm{gen},bc}$.
\end{lemma}

\begin{proof}
    The proof of the consistency of the variance estimator is similar to the proof of Theorem 3 in \citet{gu2023regression}. The proof for $\hat\sigma^2_{bc, HY}$ stays the same. We will focus on $\sigma^2_{r_{\rm{gen}}(b)}$ and the other two terms.

    The consistency of the estimator can be implied by 
    \[
        \frac{1}{n_{[k]a}}\sum\limits_{i=1}^n I(A_i=a,B_i=k)(\hat r_i - \bar\hat r_i)^2 \stackrel{P}\rightarrow \mathrm{var}(r_i|B_i=k)
    \] for $r_i = Y_i - X_i^{\rm T}\beta_{[k]}(a)$ and $X_i^{\rm T} \{\beta_{[k]}(b) - \beta_{[k]}(c)\}$, and
    \begin{align*}
        \frac{1}{n_{[k]a}}\sum\limits_{i=1}^n I(A_i=a,B_i=k)\{\beta_{[k]}(b) - \beta_{[k]}(c)\} (X_i - \bar X_{[k]a})^{\rm T}(Y_i - \bar Y_{[k]a}) \\
        \stackrel{P}\rightarrow \{\beta_{[k]}(b) - \beta_{[k]}(c)\}\Sigma_{[k]XY(a)}.
    \end{align*}
By Lemma \ref{lem6} below, we have 
\begin{align*}
    \frac{1}{n_{[k]a}} \sum\limits_{i=1}^n I(A_i=a,B_i=k)(Y_i - \bar Y_{[k]a})^2 \stackrel{P}\rightarrow \mathrm{var}\{Y_i(a)|B_i=k\},\\
    \frac{1}{n_{[k]a}} \sum\limits_{i=1}^n I(A_i=a,B_i=k)(r_i - \bar r_{[k]a})^2 \stackrel{P}\rightarrow \mathrm{var}\{r_i(a)|B_i=k\}.
\end{align*}
    Then we have 
    \begin{align*}
        \frac{1}{n_{[k]a}}\sum\limits_{i=1}^n I(A_i=a,B_i=k)(\hat r_i - \bar\hat r_i)^2 &= \frac{1}{n_{[k]a}}\sum\limits_{i=1}^n I(A_i=a,B_i=k)(r_i - \bar r_i)^2 \\
        &-\frac{2}{n_{[k]a}}\sum\limits_{i=1}^n I(A_i=a,B_i=k)(r_i - \bar r_i)(X_i-\bar X_{[k]a})^{\rm T} (\hat \beta_{[k]}(a) - \beta_{[k]}(a)) \\
        &+\frac{1}{n_{[k]a}}\sum\limits_{i=1}^n I(A_i=a,B_i=k)\big[(X_i -\bar X_{[k]a})^{\rm T}\{\hat\beta_{[k]}(a) - \beta_{[k]}(a)\}\big]^2 \\
       & = \frac{1}{n_{[k]a}}\sum\limits_{i=1}^n I(A_i=a,B_i=k)(r_i - \bar r_i)^2 + o_p(1) \\
       & \stackrel{P}\rightarrow \mathrm{var}(Y_i(a)|B_i=k).
    \end{align*}
    The third term converges to zero due to the condition that
    \[\frac{1}{n_{[k]}}\sum\limits_{i\in [k]} [X_i^{\rm T}\{\hat\beta_{[k]}(a) - \beta_{[k]}(a)\}]^2 = o_p(1)\]
    and the second term converges to zero due to the Cauchy-Schwarz inequality.
    Similarly, in the last term of the variance estimator, we can only deal with
    \[
        \big\{\hat{\beta}_{[k]}(b)-\hat{\beta}_{[k]}(c)\big\}^{\mathrm T}\left\{S_{[k]XY(b)}-S_{[k]XY(c)}\right\}.
    \]
    Also, it is sufficient to show that for any treatment $b$ and $c$, we have
    \[\{\hat{\beta}_{[k]}(b)\}^{\rm T} S_{[k]XY(c)} \stackrel{P}\rightarrow \beta^{\rm T}_{[k]}(b) \Sigma_{[k]XY(c)}.\]
    The left-hand side can be expressed as
    \begin{align*}
        \{\hat{\beta}_{[k]}(b)\}^{\rm T} S_{[k]XY(c)} &=\beta^{\rm T}_{[k]}(b) \frac{1}{n_{[k]c}} \sum\limits_{i=1}^n I(A_i=c,B_i=k) (X_i-\bar X_{[k]c})^{\rm T}(Y_i - \bar Y_{[k]c}) \\
        &+ \frac{1}{n_{[k]c}} \{\hat{\beta}_{[k]}(b)-\beta_{[k]}(b)\}^{\rm T} \sum\limits_{i=1}^n I(A_i=c,B_i=k) (X_i-\bar X_{[k]c})(Y_i - \bar Y_{[k]c}).
    \end{align*}
    As 
    \begin{align*}
        \frac{1}{n_{[k]c}} \sum\limits_{i=1}^n I(A_i=c,B_i=k) &[(X_i-\bar X_{[k]c})^{\rm T}\{\hat{\beta}_{[k]}(b)-\beta_{[k]}(b)\}]^2 \\
        &\leq \frac{1}{n_{[k]c}} \sum\limits_{i=1}^n I(B_i=k) [(X_i-\bar X_{[k]c})^{\rm T}\{\hat{\beta}_{[k]}(b)-\beta_{[k]}(b)\}]^2 \\
        &\leq \frac{1}{n_{[k]}} \sum\limits_{i=1}^n I(B_i=k) [X_i^{\rm T}\{\hat{\beta}_{[k]}(b)-\beta_{[k]}(b)\}]^2 \\
       & = o_p(1),       
    \end{align*}
    and the second term is $o_p(1)$ due to the Cauchy-Schwarz inequality, we show the convergence of the last term in the estimator. 

    Finally, we need to show that 
    \begin{align*}
        \big\{\hat{\beta}_{[k]}(b)-\hat{\beta}_{[k]}(c)\big\}^{\mathrm T} S_{[k]XX}\big\{\hat{\beta}_{[k]}(b)-\hat{\beta}_{[k]}(c)\big\} \\
        \stackrel{P}\rightarrow \big\{{\beta}_{[k]}(b)-{\beta}_{[k]}(c)\big\}^{\mathrm T} \Sigma_{[k]XX}\big\{{\beta}_{[k]}(b)-{\beta}_{[k]}(c)\big\}.
    \end{align*}
    We can write the LHS as 
    \[\frac{1}{n_{[k]a}}\sum\limits_{i=1}^n I(B_i=k)\big[(X_i - \bar X_{[k]})^{\rm T}\big\{\hat{\beta}_{[k]}(b)-\hat{\beta}_{[k]}(c)\big\}\big]^2\]
    and split the term into three parts: 
    \begin{multline*}
        \frac{1}{n_{[k]a}}\sum\limits_{i=1}^n I(B_i=k)\big[(X_i - \bar X_{[k]})^{\rm T}\big\{{\beta}_{[k]}(b)-{\beta}_{[k]}(c)\big\}\big]^2, \\
        \frac{1}{n_{[k]a}}\sum\limits_{i=1}^n I(B_i=k)\big[(X_i - \bar X_{[k]})^{\rm T}\big\{\hat{\beta}_{[k]}(b) -{\beta}_{[k]}(b) -(\hat{\beta}_{[k]}(c)-{\beta}_{[k]}(c))\big\}\big]^2
    \end{multline*}
     and their production.
     Similar to the former arguments, the first term converges to the result. The second term converges to zero due to the condition and the Cauchy-Schwarz inequality.
\end{proof}

\begin{lemma}\label{lem6}
 For $f_i = f(Y_i(0),\dots ,Y_i(A),B_i,X_i)$ which satisfies $E[f_i^2] < \infty$, we have
 \begin{align*}
        \frac{1}{n} \sum\limits_{i=1}^n f_i I(A_i = a) &\stackrel{P}\rightarrow \sum\limits_{k \in \mathcal{K}} p_{[k]} \pi_{[k]a} E(f_i|B_i=k), \\
        \frac{1}{n} \sum\limits_{i=1}^n f_i I(A_i = a,B_i=k) &\stackrel{P}\rightarrow  p_{[k]} \pi_{[k]a} E(f_i|B_i=k).
    \end{align*}
\end{lemma}
\begin{proof}
    It is a generalized version of Lemma C.4 from \cite{Bugni2019} and Lemma 1 from \cite{Ma2020regression}. They proved their lemmas by setting $f_i = f(Y_i(0),\dots,Y_i(A),B_i)$ and $f_i = f(Y_i(0),Y_i(1),X_i)$. The proof can be easily generalized to $f_i = f(Y_i(0),\dots ,Y_i(A),B_i,X_i)$ so we omit it. 
\end{proof}

\section{Additional simulations}
\begin{figure}[htbp]
\begin{subfigure}{1\textwidth}
    \centering
    \caption{}\label{figS1a}
    \includegraphics[width=0.87\linewidth,height=0.38\linewidth]{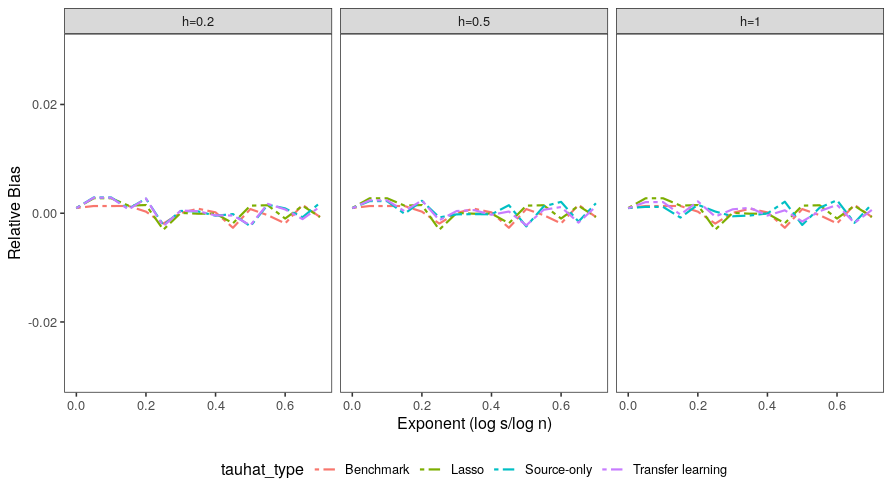}
 
\end{subfigure}

\begin{subfigure}{1\textwidth}
    \centering
    \caption{}\label{figS1b}
    \includegraphics[width=0.85\linewidth,height=0.38\linewidth]{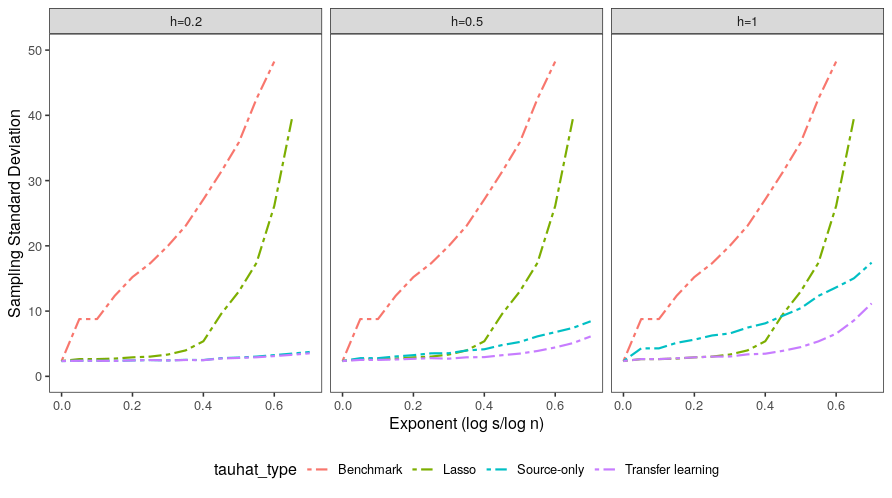}
\end{subfigure}
\begin{subfigure}{1\textwidth}
    \centering
    \caption{}\label{figS1c}
    \includegraphics[width=0.85\linewidth,height=0.38\linewidth]{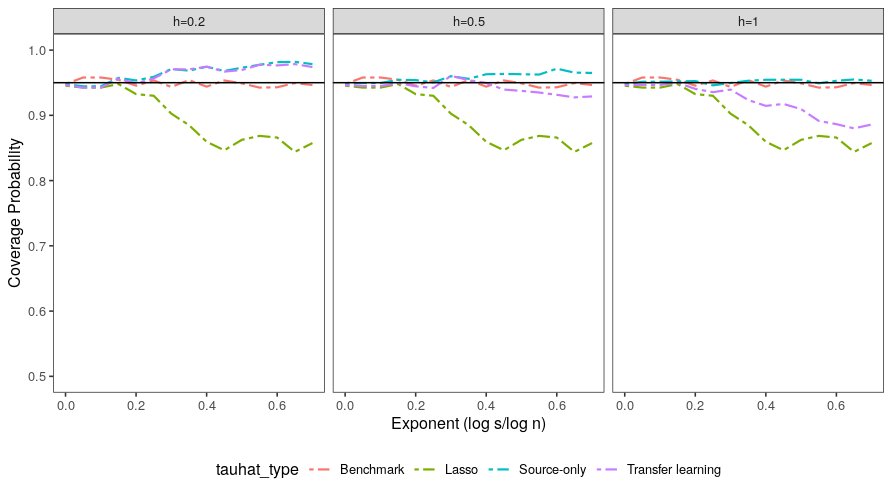}
\end{subfigure}
\caption{Simulations in Case 1 under stratified block randomization with unequal allocation ratios. Each column represents different $h$ in the simulation settings, resulting in different $l_1$ norms of the bias. (a) Relative bias for $\hat\tau_{\rm ben}$, $\hat\tau_{\rm lasso}$, $\hat\tau_{\rm so}$ and $\hat\tau_{\rm tl}$. (b) Sampling standard deviation for$\hat\tau_{\rm ben}$, $\hat\tau_{\rm lasso}$, $\hat\tau_{\rm so}$ and $\hat\tau_{\rm tl}$. (c) Coverage probabilities for $\hat\tau_{\rm ben}$, $\hat\tau_{\rm lasso}$, $\hat\tau_{\rm so}$ and $\hat\tau_{\rm tl}$.}\label{figS1}
 \end{figure}

\begin{figure}[htbp]
\begin{subfigure}{1\textwidth}
    \centering
        \caption{}
    \includegraphics[width=0.87\linewidth,height=0.38\linewidth]{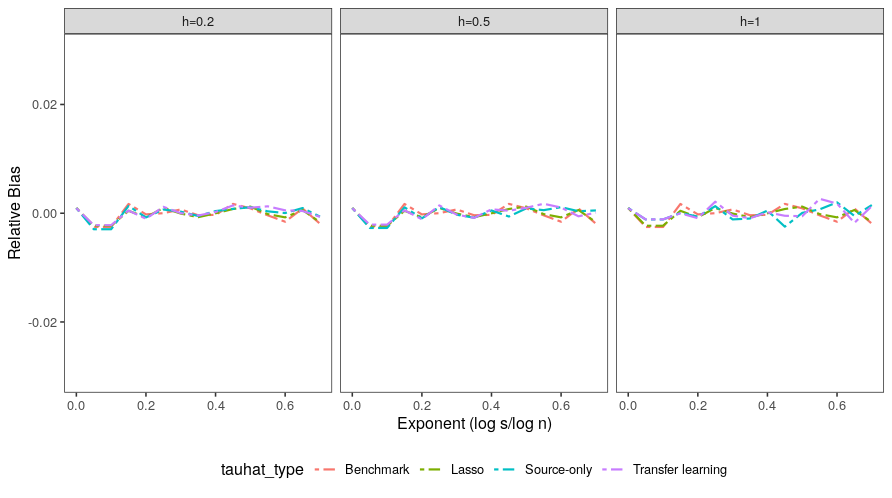}

\end{subfigure}

\begin{subfigure}{1\textwidth}
    \centering
        \caption{}
    \includegraphics[width=0.85\linewidth,height=0.38\linewidth]{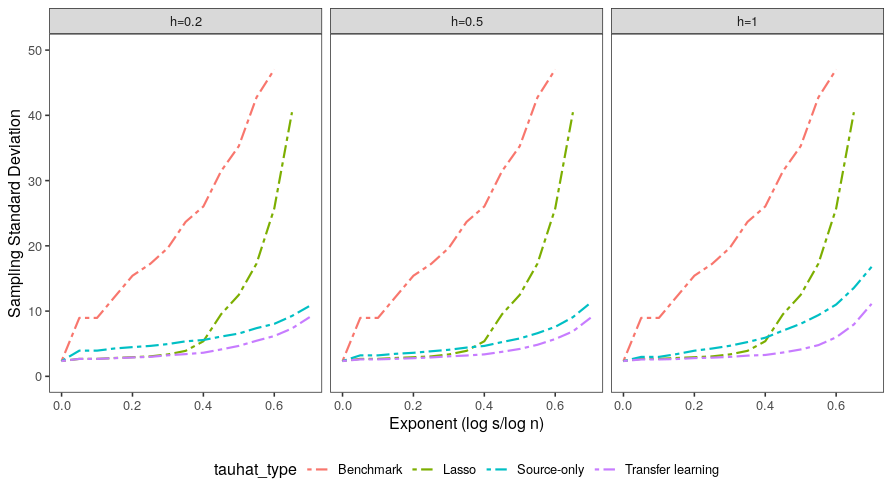}

\end{subfigure}
\begin{subfigure}{1\textwidth}
    \centering
        \caption{}
    \includegraphics[width=0.85\linewidth,height=0.38\linewidth]{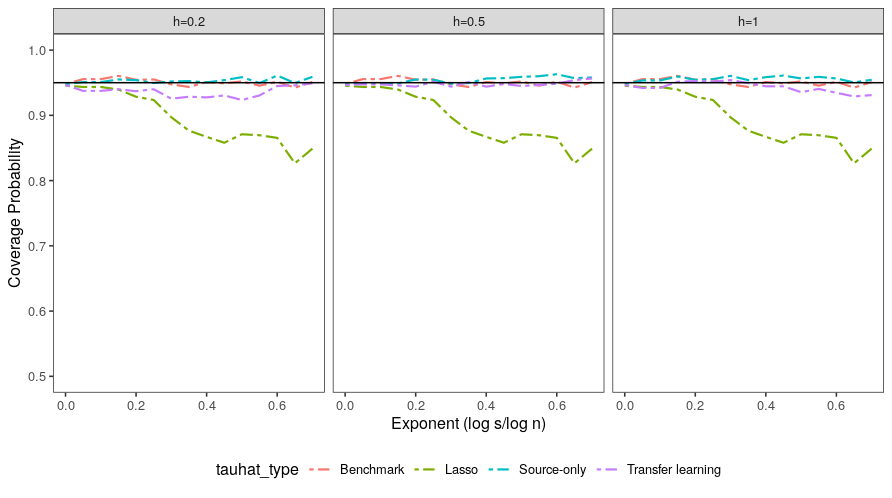}

\end{subfigure}
\caption{Simulations in Case 2 under stratified block randomization with unequal allocation ratios. Each column represents different $h$ in the simulation settings, resulting in different $l_1$ norms of the bias. (a) Relative bias for $\hat\tau_{\rm ben}$, $\hat\tau_{\rm lasso}$, $\hat\tau_{\rm so}$ and $\hat\tau_{\rm tl}$. (b) Sampling standard deviation for$\hat\tau_{\rm ben}$, $\hat\tau_{\rm lasso}$, $\hat\tau_{\rm so}$ and $\hat\tau_{\rm tl}$. (c) Coverage probabilities for $\hat\tau_{\rm ben}$, $\hat\tau_{\rm lasso}$, $\hat\tau_{\rm so}$ and $\hat\tau_{\rm tl}$.}\label{figS2}
 \end{figure}

\begin{figure}[htbp]
\begin{subfigure}{1\textwidth}
    \centering
        \caption{}
    \includegraphics[width=0.87\linewidth,height=0.38\linewidth]{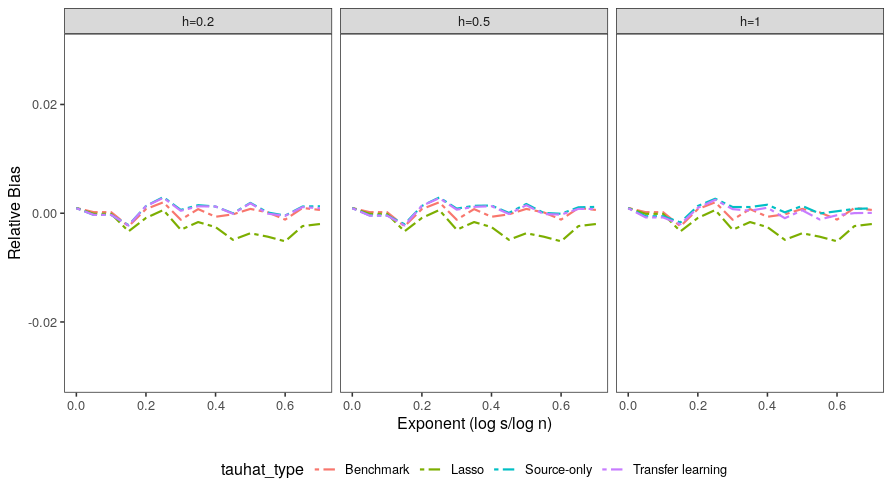}

\end{subfigure}

\begin{subfigure}{1\textwidth}
    \centering
        \caption{}
    \includegraphics[width=0.85\linewidth,height=0.38\linewidth]{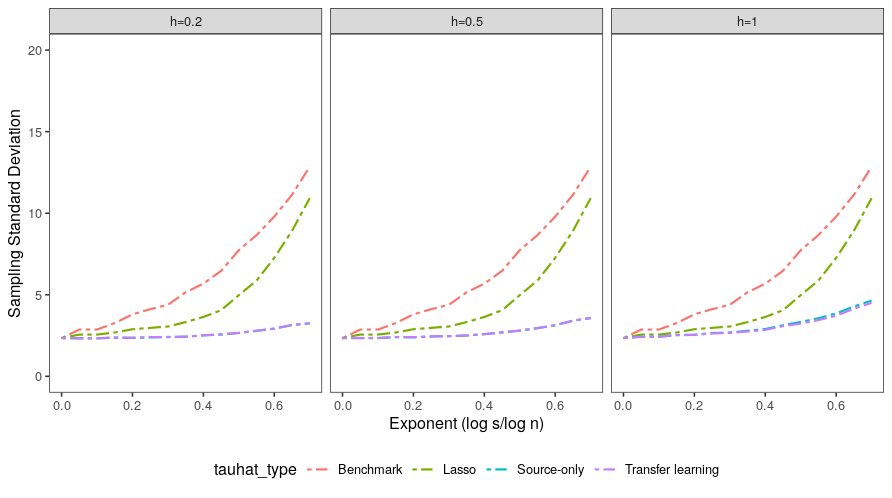}

\end{subfigure}
\begin{subfigure}{1\textwidth}
    \centering
        \caption{}
    \includegraphics[width=0.85\linewidth,height=0.38\linewidth]{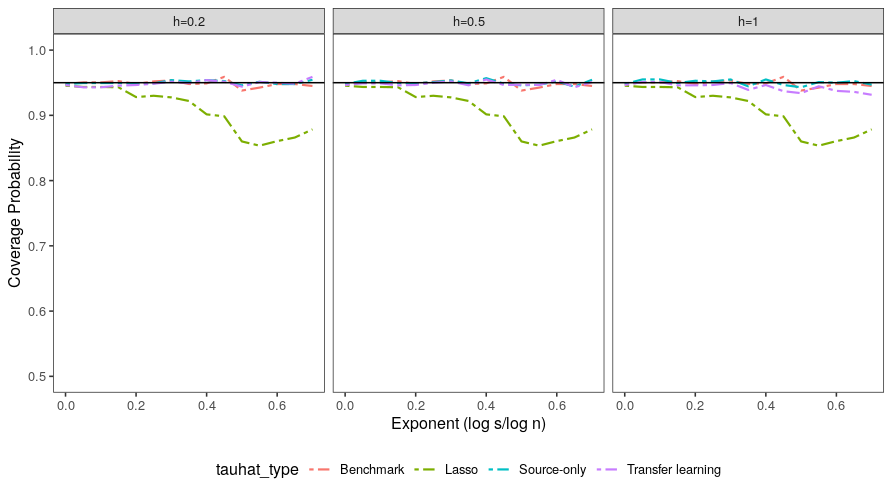}

\end{subfigure}
\caption{Simulations in Case 3 under stratified block randomization with unequal allocation ratios. Each column represents different $h$ in the simulation settings, resulting in different $l_1$ norms of the bias. (a) Relative bias for $\hat\tau_{\rm ben}$, $\hat\tau_{\rm lasso}$, $\hat\tau_{\rm so}$ and $\hat\tau_{\rm tl}$. (b) Sampling standard deviation for$\hat\tau_{\rm ben}$, $\hat\tau_{\rm lasso}$, $\hat\tau_{\rm so}$ and $\hat\tau_{\rm tl}$. (c) Coverage probabilities for $\hat\tau_{\rm ben}$, $\hat\tau_{\rm lasso}$, $\hat\tau_{\rm so}$ and $\hat\tau_{\rm tl}$.}\label{figS3}
 \end{figure}

Figures \ref{figS1}--\ref{figS3} represent the simulation results when the allocation ratios are unequal within each stratum in the main text. The allocation ratios are 1:1:1 in the first strata and 2:2:1 in the other strata, and the block size for the stratified block randomization is 15. Other settings are the same as the settings in Section \ref{sec7} in the main text. The simulation results are similar to the results in the main text.

\end{document}